\documentclass[11pt,a4]{article}
\usepackage{color}
\usepackage{graphicx}  
\usepackage[all]{xypic}
\usepackage{psfrag}
\usepackage{tocbibind} 
\usepackage{epsf}
\usepackage{epic}
\usepackage{eepic}
\usepackage{epsfig}
\usepackage{wrapfig}
\usepackage{a4}
\usepackage{amssymb,latexsym}
\usepackage[mathscr]{eucal}
\usepackage{citesort}
\usepackage[active]{srcltx}
\usepackage{url}
\usepackage{fancybox}
\usepackage{ntheorem}
\usepackage{mhequ}
\usepackage[g]{esvect}
\usepackage{epstopdf}

\newcommand{\qedskip}{\hfill $\Box$\medskip}

\newcommand{\syncx}[1]{}

\newcommand{\restrict}[1]{\ptcr{here starts a restrict environment, not to be displayed in the arxiv version}{\color{blue}{ #1}}
 \ptcr{here ends a restrict environment}}

\newcommand{\citex}[3]{[#1#2]\cite{ChCausalityv1}}

\newcommand{\citexG}[3]{[#1#2]\cite{ChGrant}\ptc{synchronize crossreference label #3
when posting, and when merging things}}

\newcommand{\ptcxx}[1]{\mnote{{\bf ptc:} {\color{red} #1}}}

\newcommand{\mnotex}[1]
{\protect{\stepcounter{mnotecount}}$^{\mbox{\footnotesize
$
\bullet$\themnotecount}}$ \marginpar{
\raggedright\tiny\em
$\!\!\!\!\!\!\,\bullet$\themnotecount: #1} }

\newcommand{\jamesx}[1]{}
\renewcommand{\jamesx}[1]{{\mnote{{\color{black}{\bf jg:}
#1} }}}

\newcommand{\h}[2]{#1\dotfill\ #2\\\ptc{fixme}}

\newcommand{\gammaXp}{\gamma_p}
\newcommand{\gammaXq}{\gamma_q}

\newcommand{\hypmco}[1]{\hyp_{\mcO;#1}{}}
\newcommand{\hyptwo}[2]{\hyp_{#1;#2}{}}

\newcommand{\gst}{>_{\!\!*}}

\newcommand{\wlqg}{W^{\ell,q}}
\newcommand{\wlqgl}{\wlqg_\loc}

\newcommand{\ol}{\overline}

\def\nz{\ifmmode {I\hskip -3pt N} \else {\hbox {$I\hskip -3pt N$}}\fi}
\def\zz{\ifmmode {Z\hskip -4.8pt Z} \else
       {\hbox {$Z\hskip -4.8pt Z$}}\fi}
\def\qz{\ifmmode {Q\hskip -5.0pt\vrule height6.0pt depth 0pt
       \hskip 6pt} \else {\hbox
       {$Q\hskip -5.0pt\vrule height6.0pt depth 0pt\hskip 6pt$}}\fi}
\def\rz{\ifmmode {I\hskip -3pt R} \else {\hbox {$I\hskip -3pt R$}}\fi}
\def\cz{\ifmmode {C\hskip -4.8pt\vrule height5.8pt\hskip 6.3pt} \else
       {\hbox {$C\hskip -4.8pt\vrule height5.8pt\hskip 6.3pt$}}\fi}
\def\au{{\setbox0=\hbox{\lower1.36775ex\hbox{''}\kern-.05em}\dp0=.36775ex\hs
kip0pt\box0}}
\def\ao{{}\kern-.10em\hbox{``}}

\newcommand\Gregbeq{\begin{eqnarray}}
\newcommand\Gregeeq{\end{eqnarray}}

\def\h1{{\hat 1}}
\def\h2{{\hat 2}}

\def\3f{\frac{3}{2}}

\def\bR{{\bar R}}

\def\cuNorma{ {\mcU}}

\newcommand{\wci}{\widetilde{\cal{I}}}
\newcommand{\ci}{{\cal{I}}}
\newcommand{\cI}{{\check{I}{}}}

\newcommand{\roscoff}[1]{}

{\catcode `\@=11 \global\let\AddToReset=\@addtoreset}
\AddToReset{figure}{section}

\DeclareFontFamily{OT1}{rsfs}{}
\DeclareFontShape{OT1}{rsfs}{m}{n}{ <-7> rsfs5 <7-10> rsfs7 <10-> rsfs10}{}
\DeclareMathAlphabet{\mycal}{OT1}{rsfs}{m}{n}

{\catcode `\@=11 \global\let\AddToReset=\@addtoreset}
\AddToReset{equation}{section}

\newcounter{mnotecount}[section]

\renewcommand{\themnotecount}{\thesection.\arabic{mnotecount}}

\newcommand{\tmcM}{\,\,\,\,\widetilde{\!\!\!\!\mcM}}%

\newcommand{\jlcax}[1]{}
\newcommand{\eean}{\nonumber\end{eqnarray}}

\newcommand{\ptcr}[1]{{\color{red}\mnote{{\color{red}{\bf ptc:}
#1} }}}

\newcommand{\kk}[1]{}

\newcommand{\beq}{\begin{equation}}

\newcommand{\FS}       
                  {F}

\newcommand{\HS} 
       {H_{\mbox{\scriptsize volume}}}

{\ptc{this should be removed in the oberwolfach version}}%

\newcommand{\eeal}[1]{\label{#1}\end{eqnarray}}
\newcommand{\id}{\textrm{id}}
\newcommand{\bed}{\begin{deqarr}}
\newcommand{\eed}{\end{deqarr}}
\newcommand{\bedl}[1]{\begin{deqarr}\label{#1}}
\newcommand{\eedl}[2]{\arrlabel{#1}\label{#2}\end{deqarr}}

\newcommand{\loc}{\textrm{\scriptsize\upshape loc}}

\newcommand{\mcO}{{\mycal O}}
\newcommand{\mcU}{{\mycal U}}

\newcommand{\bel}[1]{\begin{equation}\label{#1}}
\newcommand{\bea}{\begin{eqnarray}}
\newcommand{\bean}{\begin{eqnarray}\nonumber}
\newcommand{\beal}[1]{\begin{eqnarray}\label{#1}}
\newcommand{\eea}{\end{eqnarray}}

\newcommand{\Eq}[1]{Equation~\eq{#1}}

\newcommand{\qed}{\hfill $\Box$ \medskip}
\newcommand{\proof}{\noindent {\sc Proof:\ }}
\newcommand{\be}{\begin{equation}}
\newcommand{\eeq}{\end{equation}}
\newcommand{\ee}{\end{equation}}
\newcommand{\beqa}{\begin{eqnarray}}
\newcommand{\eeqa}{\end{eqnarray}}
\newcommand{\beqan}{\begin{eqnarray*}}
\newcommand{\eeqan}{\end{eqnarray*}}
\newcommand{\ba}{\begin{array}}
\newcommand{\ea}{\end{array}}

\newcommand{\hyp}{\mycal S}

\newcommand{\mcM}{{\mycal M}}
\newcommand{\mcD}{{\mycal D}}

\newcommand{\mcV}{{\mycal V}}

\newcommand{\calW}{{\cal W}}
\newcommand{\cU}{{\cal U}}
\newcommand{\cV}{{\cal V}}

\newcommand{\mnote}[1]
{\protect{\stepcounter{mnotecount}}$^{\mbox{\footnotesize
$
\bullet$\themnotecount}}$ \marginpar{
\raggedright\tiny\em
$\!\!\!\!\!\!\,\bullet$\themnotecount: #1} }

\newcommand{\warn}[1]
{\protect{\stepcounter{mnotecount}}$^{\mbox{\footnotesize
$
\bullet$\themnotecount}}$ \marginpar{
\raggedright\tiny\em
$\!\!\!\!\!\!\,\bullet$\themnotecount: {\bf Warning:} #1} }

\newcommand{\R}{\mathbb R}

\newcommand{\N}{\mathbb N}

\newcommand{\eq}[1]{(\ref{#1})}

\newcommand{\ptc}[1]{\mnote{{\bf ptc:}#1}}

\newcommand{\beqar}{\begin{deqarr}}
\newcommand{\eeqar}{\end{deqarr}}

\newcommand{\beaa}{\begin{eqnarray*}}
\newcommand{\eeaa}{\end{eqnarray*}}

\newcommand{\tr}{\mbox{tr}}

\newcommand{\hg}{{\hat g}}

\newcommand{\myspaceloc}{{\mathrm{space,loc}}}

\newcommand{\Hssp}{H^s_{\myspaceloc}}

\newcommand{\Hksp}{H^{k}_{\myspaceloc}}

\newcommand{\Hspsp}{H^{s+1}_{\myspaceloc}}
\newcommand{\Hsmsp}{H^{s-1}_{\myspaceloc}}

\newcommand{\Wkpsp}{W^{k,p}_{\myspaceloc}}
\newcommand{\Wkppsp}{W^{k+1,p}_{\myspaceloc}}
\newcommand{\Wkmpsp}{W^{k-m,p}_{\myspaceloc}}
\newcommand{\Wlmqsp}{W^{\ell-m,q}_{\myspaceloc}}
\newcommand{\Wlqsp}{W^{\ell,q}_{\myspaceloc}}

\newcommand{\Wujvjsp}{W^{u_m+j ,v_j}_{\myspaceloc}}
\newcommand{\Wujvjmsp}{W^{u_m+j ,v_{j-1}}_{\myspaceloc}}
\newcommand{\Wujmvjsp}{W^{u_m+j-1 ,v_j}_{\myspaceloc}}
\newcommand{\Wujpvjsp}{W^{u_m+j+1 ,v_j}_{\myspaceloc}}
\newcommand{\Wujpvjpsp}{W^{u_m+j+1 ,v_{j+1}}_{\myspaceloc}}

\newcommand{\tmg}{(\tilde M,\tilde\sptmetric,\tilde\imath)}
\newcommand{\tmgo}{(\tilde M_1,\tilde\sptmetric_1,\tilde \imath_1)}
\newcommand{\mw}{{\cal M}_{W,\hyp}}
\newcommand{\mwm}{{\cal M}_W(M,\sptmetric,i)}

\newcommand{\sptmetric}{g}
\newcommand{\bethm}{\begin{Theorem}}
\newcommand{\et}{\end{Theorem}}
\newcommand{\bl}{\begin{Lemma}}
\newcommand{\veps}{{\hyp}}
\newcommand{\phie}{\emptyset}

\newcommand{\cv}{{\cal{V}}}
\newcommand{\cy}{{\cal{Y}}}
\newcommand{\cw}{{\cal{W}}}

\renewcommand{\restrict}[1]{\ptcxx{here starts a restrict environment}}

\renewcommand{\roscoff}[1]{#1}

\renewcommand{\ptcxx}[1]{}

\renewcommand{\citexG}[3]{[#1#2]\cite{ChGrant}}

\renewcommand{\jamesx}[1]{{\mnote{{\color{black}{\bf jg:}
#1} }}}

\newtheorem{Theorem} {\sc  Theorem\rm} [section]
\newtheorem{Corollary} [Theorem] {\sc  Corollary\rm}
\newtheorem{Lemma} [Theorem] {\sc  Lemma\rm}
\newtheorem{Proposition} [Theorem] {\sc  Proposition\rm}

\newtheorem{Definition}[Theorem]{\sc  Definition\rm}

\theorembodyfont{\upshape}
\newtheorem{Remark}[Theorem]{\sc  Remark\rm}
\newtheorem{remark}[Theorem]{\sc  Remark\rm}
\newtheorem{Remarks}[Theorem]{\sc  Remarks\rm}

\theoremsymbol{\ensuremath{\diamondsuit}}
\newcommand{\fcoco}{\small}
\theorembodyfont{\fcoco}\theoremseparator{}\theoremindent0.5cm

\theoremstyle{nonumberplain}\theorembodyfont{\fcoco}
\theoremseparator{}\theoremindent0.5cm
\newtheorem{coco}{}

\DeclareFontFamily{OT1}{rsfs}{}
\DeclareFontShape{OT1}{rsfs}{m}{n}{ <-7> rsfs5 <7-10> rsfs7 <10-> rsfs10}{}
\DeclareMathAlphabet{\mycal}{OT1}{rsfs}{m}{n}

\begin{document}
\title{On maximal globally hyperbolic vacuum space-times\thanks{Preprint UWThPh-2011-44}}
\author{Piotr T.\ Chru\'{s}ciel\thanks{I.H.E.S,
Bures-sur Yvette, and Gravitational Physics, University of
Vienna. {URL \protect\url{http://homepage.univie.ac.at/piotr.chrusciel}}}
{}\thanks{The author is grateful to l'IHES for hospitality and support during part of the work on this paper.  He was
also supported in part by the Polish Ministry of Science and
Higher Education grant Nr N N201 372736.}
}
\maketitle
\abstract{We prove existence and uniqueness of maximal global hyperbolic developments of vacuum general relativistic initial data sets with initial data $(g,K)$ in Sobolev spaces $H^s\oplus H^{s-1}$, $\N\ni s > n/2 +1$.}
\maketitle

\tableofcontents

\section{Introduction}

One of the many contributions of Yvonne Choquet-Bruhat to the foundations of mathematical general relativity is the proof of the local well-posedness of the initial value problem for the Einstein equations~\cite{ChBActa}, followed by
the celebrated Choquet-Bruhat -- Geroch theorem~\cite{ChoquetBruhatGeroch69}, which asserts that to every smooth vacuum general relativistic initial data set $(\hyp,g,K)$ one can associate a
unique, up to isometries, smooth solution of the vacuum Einstein equations. Now, there is a gap between the smoothness assumed in this last result and the classical local existence theory~\cite{hughes:kato:marsden}, where solutions with Sobolev initial data $(\gamma,K)\in H^{s}\oplus H^{s-1}$
are constructed for $s>n/2+1$. The aim of this work is to make a step towards bridging this gap. It is a great pleasure to dedicate the following theorem to Yvonne Choquet-Bruhat for her ninetieth birthday: 

 \syncx{file Completeness.tex}
\newcommand{\hcV}{\widehat{\cV}}
\newcommand{\omcM}{\,\,\,\,\overline{\!\!\!\!\mcM}}
\newcommand{\Azero}{A0}
\newcommand{\Aone}{A1}
\newcommand{\Aonep}{A1'}
\newcommand{\Hone}{H1}
\newcommand{\Htwohere}{H2}
\newcommand{\Atwo}{A2}
\newcommand{\Athree}{A3}
\newcommand{\cohere}{\mcO}
\newcommand{\spaceloc}{{\mathrm{space,loc}}}

{\syncx{some introductory comments needed, perhaps from runCompleteness.tex from James Grant's directory}{}}

\begin{Theorem}
 \label{T11XI11.1}
Consider  a vacuum Cauchy data set
$(\hyp ,\gamma,K)$,
where $\hyp$ is an $n$-dimensional manifold, $\gamma \in H^{s}_\loc(\hyp)$ is a
Riemannian metric on $\hyp $, and $K \in H^{s-1}_\loc(\hyp)$ is a symmetric two--tensor on
$\hyp $, satisfying the general relativistic vacuum constraint equations, where  $\N\ni s>n/2+1$, $n\ge 3$.
Then there exists a Lorentzian manifold $(\mcM,\sptmetric)$ with a  $\Hssp$-metric, {\em unique up to isometries} within the  $\Hssp$ class,  {\em inextendible} in the class of
globally hyperbolic space--times with a $\Hssp$ vacuum metric and with an embedding $i:\hyp \rightarrow \mcM$ such
that
$i^*g=\gamma$, and such that $K$ corresponds to the extrinsic curvature tensor of
$i(\hyp )$ in $\mcM$.
\end{Theorem}

We have considered the vacuum Einstein equations for definiteness, similar results can be established for Einstein equations coupled with appropriate matter sources, e.g. for the Einstein-minimally coupled scalar field equations, or the Einstein-Yang-Mills-Higgs equations, or in fact for any quasi-linear systems of wave-equations in space-dimensions $n\ge 1$.

The manifold $(\mcM,g)$ of the theorem is called the {\em maximal globally hyperbolic vacuum development of
$(\hyp ,\gamma,K)$}; it is yet another classical result of Yvonne Choquet-Bruhat that $(\mcM,g)$ is independent of $s$ for $s>n/2+1$.
 \syncx{prove independence of s,  for $s\in\N$? $s\in \R$?}

To avoid ambiguities, global hyperbolicity here is the requirement that every inextendible causal curve meets $i(\hyp)$ precisely once.

There is little doubt that the condition  $\N\ni s>n/2+1$ can be relaxed to $\R\ni s>n/2+1$ using paradifferential techniques, see e.g.~\cite{AlinhacGerard,MetivierParaDiff}. We reduce this question to the problem of verifying conditions \Hone-\Athree, p.~\pageref{H7XI11.H}, compare Theorem~\ref{T7XI11.1} below.
It is conceivable that the result  generalises to more general classes of initial data for which local existence and uniqueness of solutions holds, such as e.g. those considered in \cite{KlainermanRodnianski:r1,%
KlainermanRodnianski:r2,KlainermanRodnianski:r3,PlanchonRodnianski,%
ParlongueCRAS,WangRough,SmithTataru}, but this remains to be seen.

{
\syncx{make a coco environment for the paragraph below, and rewordings needed}{}}
The proof here is an adaptation of that in~\cite{Chorbits},
using the Planchon-Rodnianski uniqueness argument~\cite{PlanchonRodnianski},%
\footnote{The reference~\cite{PlanchonRodnianski} is not available at the time of publication of this work, but the relevant argument can be found in~\cite[Section~4.3]{CGP}.}
an extension of the analysis in~\cite[Appendix~A]{BartnikChrusciel1} to Lorentzian manifolds with  $\Hssp$ metrics,
and the causality theory for continuous metrics in~\cite{ChGrant}. In fact, this last work was carried out with the current problem in mind.

It might be useful to comment upon the differentiability thresholds that arise
in previous proofs of the theorem.
First, all previous proofs use various elements of causality theory which have only
been consistently developed using standard approaches for smooth, or $C^2$~\cite{ChCausalityv1} metrics. So, without further detailed justifications, that part of the proof that appeals to causality theory would require  at least $C^2$-differentiability
of the metric.
The original proof in~\cite{ChoquetBruhatGeroch69} assumes explicitly
smoothness at the outset, and invokes existence and uniqueness of geodesics,
which fails for metrics which are not $C^{1,1}$. Similarly geodesics are invoked
in the proofs given in~\cite{HE,RingstroemBook,CBY}.

This author's  argument  in~\cite{Chorbits} (which proves a more general result concerning orbits of Killing vectors, with the Choquet-Bruhat -- Geroch theorem in the smooth category of metrics being a straightforward consequence of Proposition~2.2 there) was presented for smooth metrics because
neither the low-differentiability causality theory, nor the
Planchon-Rodnianski uniqueness argument~\cite{PlanchonRodnianski} were
available at the time. As such, the proof in~\cite{Chorbits} was written using arguments which were expected to  generalise to metrics with Sobolev differentiability if the
associated causality theory goes through. Inspection of~\cite{Chorbits} shows
that the elements of causality theory needed there arise in the proof of Lemma~2.3 of that reference, and are
\begin{enumerate}
 \item[(1)] existence and causality of accumulation curves;
 \item[(2)] the fact that a causal curve which
is not null everywhere can be deformed to a timelike curve at end points fixed; and
 \item[(3)]
 existence of null geodesic generators of  $\partial \mcD_J$.
\end{enumerate}
It is not obvious,
but proved in \cite{ChGrant} (see also~\cite{FathiSiconolfiTime}), that point (1) remains true for continuous metrics,
but that  point (2) is wrong for continuous metrics; then the usual arguments addressing (3) fail.
This part of the argument is replaced by the rather more involved argument
starting after the proof of Lemma~\ref{L21IX11.1}, p.~\pageref{L21IX11.1}
below.

\section{Existence of maximal developments}
\label{contProofs}

As a first step in the proof of the Choquet-Bruhat -- Geroch theorem, one
constructs space-times which are maximal with respect to a set of properties.
 \syncx{how do I handle black hole extensions for this? and adS type cauchy problems? and characteristic cauchy problems? it would make it worthwile to add those}
This begs the question, if and when is such a construction possible. We start by addressing this. Some notation is in order.

Let $W$ denote a set of properties of a manifold, possibly equipped with some supplementary structure such as a metric.
Here all manifolds are connected, paracompact,
Hausdorff,  of at least $C^1$-differentiability class. When talking about
space-time, the dimension will be denoted by $n+1$. Thus,
spacelike hypersurfaces, or their models, will be of dimension $n$.
The property $W$ will include differentiability requirements,
{\em e.g.\ }$C^{k,\alpha}$, or analyticity, or some Sobolev class, and it might, or might-not, include some further requirements.

A manifold will be said to be Lorentzian if it is equipped with a metric tensor,
perhaps defined only almost everywhere, of a differentiability class
adapted to $W$. For example,
a natural class $W$ could be manifolds with a  $C^{k,\alpha}$ atlas, $k\ge 1$, and metrics of  $C^{k-1,\alpha}$-differentiability class. It is useful to
keep in mind that $W$ can denote a rather complicated structure. For  the purpose of the Cauchy problem in general relativity we will be using a $\Hssp$-structure, defined as follows:

\begin{Definition}
 \label{D17IX11.1}
 Let $s\in\R$.
 A Lorentzian manifold $(\mcM,g)$ will be said to have a metric of $\Hssp$-differentiability class if every point $p\in \mcM$ has a coordinate neighborhood $\mcU_p=I\times \mcV_p$, where $I\subset\R$ is the range of a local time coordinate $t\equiv x^0$, with the following properties:
 On every level set $\hyp_\tau\subset \mcU_p$ of the coordinate $t$ the metric components $g_{\mu\nu}$ are of $H^s$-differentiability class, and their time-derivatives of order $0\le k\le \lfloor s \rfloor$ are of $H^{s-k}$-differentiability. Furthermore the functions
$$
 I \ni t \mapsto \|\partial^k_0 g_{\mu\nu} |_{\hyp_t}\|_{H^{s-k}}
  \;,
 \qquad 0\le k \le  \lfloor s \rfloor
 \;,
$$
 are continuous.
\end{Definition}

Thus, the subscript ``space" in $\Hssp$ denotes the fact that the differentiability of the metric is defined in terms of Sobolev spaces on spacelike hypersurfaces. The subscript ``loc", shorthand for ``local", refers to the fact that  the integrals defining the $H^s$ norms are finite on conditionally compact sets, but not necessarily so for sets with non-compact closure.

Our results show  that the definition is well suited for the problem at hand when $s\in \N$, $s>n/2+1$. It is likely that it will require fine-tuning for other values of $s$.

One expects that the maximal atlas compatible with a $\Hssp$ structure has transition maps which are of differentiability class $\Hspsp$. A version of this, which constitutes one of the elements of the proof of Theorem~\ref{T11XI11.1}, is established in Proposition~\ref{PL5XI11.1}  below for $s\in \N$ satisfying $s>n/2+1$.

Since all our manifolds are assumed to be $C^1$, maps between them will also be $C^1$ in any case, unless explicitly indicated otherwise.

A Lorentzian manifold will be called vacuum if  the equations
$R_{\mu\nu}=0$  can be defined, perhaps in a
distributional sense, and if $R_{\mu\nu}=0$ holds. Note that the Christoffel symbols can be defined for metrics with $g^{\mu\nu}\in L^\infty_\loc$ and which have distributional derivatives in $L^1_\loc$. The equation $R_{\mu\nu}=0$ can be defined in a distributional sense if moreover the distributional derivatives of the metric are in $L^2_\loc$.

The standard theory of PDEs leads to $\Hssp$-solutions of the Einstein equations with $\R\ni s> n/2+1$, with an atlas in which the coordinate functions are harmonic~\cite{hughes:kato:marsden,PlanchonRodnianski} (compare~\cite[Section~4.3]{CGP}).

The work in~\cite{KlainermanRodnianski:r1}
together with \cite[Theorem~7.1]{Maxwell:Rough}
establishes local existence of vacuum metrics in dimension $3+1$ for asymptotically flat initial data on
$
\R^3$ with $\tr K =0$ and $(\gamma,K)\in  H^{s }_\loc\times  H^{s-1}_\loc$ with $s>2$; see also~\cite{Choquet-Bruhat:safari}. The more recent results of~\cite{KRS} are expected to lower this threshold to $s=2$.

As we will be solving the Cauchy problem, we will need to consider an embedding $i$ of a spacelike hypersurface $\hyp$ into $(\mcM,g)$. The embedding should be compatible with the available structures; we will  say that $i$ is \emph{$W$-compatible} when this is the case. For example, for $C^{k,\alpha}$, or smooth, or analytic, manifolds it is natural to consider maps which are also of $C^{k,\alpha}$ class, or smooth, or analytic. So, in this case, a $W$-compatible embedding is required to be $C^{k,\alpha}$, or smooth, etc.  For manifolds with $\Hssp$ metrics it is natural to consider embeddings $i:\hyp\to \mcM$ such that the pull-backs $(\gamma,K)$ of the metric and of the extrinsic curvature from $i(\hyp)$ to $\hyp$ are in $H^s_\loc\otimes H^{s-1}_\loc$; this is our definition of $W$-compatible embedding for $\Hssp$ manifolds. The resulting hypersurfaces $i(\hyp)$ will be called $W$-compatible.

A reader only interested in smooth vacuum space-times can assume that $W$ is the property that $(\mcM,g)$ is  a smooth, Hausdorff, paracompact, connected globally hyperbolic vacuum  manifold with a smooth metric. A $W$-compatible embedding $i$ then means that $i$ is smooth and that $i(\hyp)$ is Cauchy,
 while a $W$--compatible submanifold means a smooth submanifold. Similarly for $C^{k,\alpha}$ or analytic manifolds.

As such, the next theorem works with any notions of $W$--manifold and $W$--compatible embedding which can be formulated within the framework of set theory as described e.g.\ in~\cite[Appendix]{KelleyTopologyReprinted}, under the following proviso:  The property $W$ will be said to be \emph{chain-inheriting} if both the manifold $\omcM$ of \eq{12IX12.1} and the associated embedding $\bar \imath$ of $\hyp$ have the property $W$ whenever all the manifolds $\tmcM$ in the union there and the embeddings $\tilde \imath$ do. An example of chain-inheriting property $W$ is ``$\mcM$ is a  $C^1$ manifold with a continuous metric of X-differentiability", where $X$ stands for $C^k$, with $k\in\N\cup\{\infty,\omega\}$, or $C^{k,\alpha}$,   or $\Hssp$, with the corresponding property for the embeddings.

To make things clear, for the purposes of Theorem~\ref{T11XI11.1}  the property $W$  in Theorem~\ref{TL12IX11.1} below is: ``$(\mcM,g)$ is a Hausdorff, paracompact, connected vacuum $C^1$ manifold with a Lorentzian metric of $\Hssp$-differentiability class".
However, we did not formulate the result in this way as Theorem~\ref{TL12IX11.1} has wider applicability, e.g.\ for the construction of extensions of black-hole exteriors.

We  have:

\begin{Theorem}
 \label{TL12IX11.1}
 Let $\hyp$ be a  $n$-dimensional manifold and
let $(\mcM,\sptmetric,i)$ be a Lorentzian $(n+1)$-dimensional $W$-manifold $(\mcM,g)$
with a $W$--compatible embedding $i:\hyp\to \mcM$.
Suppose that the property $W$ is chain-inheriting, and that
\bean
 && \mbox{the only isometry of a $W$-manifold $(\mcM,\sptmetric)$ which is the identity }
\\
 &&\mbox{on  a $W$-compatible
 hypersurface
  is the identity map.}
\eeal{17IX11.1}
Then there exists a
Lorentzian  $W$-manifold
$( \tmcM ,\tilde \sptmetric,\tilde \imath )$ with a $W$-compatible embedding
$\tilde \imath :\hyp\to \tmcM $  and a $C^1$ isometric
embedding $\Phi:\mcM\rightarrow \tmcM $ satisfying $\tilde \imath  = \Phi\circ i$ such that
$\tmcM $ is inextendible in the class of  Lorentzian $W$-manifolds with a
$W$-compatible embedding of $\hyp$.
\end{Theorem}

\begin{Remarks}
\begin{enumerate}
\item
The $C^1$-differentiability threshold for $\hyp$ and $\mcM$ cannot be weakened in
the proof below, since we are using the Whitney embedding theorem to turn the collection of manifolds into a set. The author ignores whether or not the $C^1$-differentiability is necessary.

\item
One expects the differentiability of $\Phi$ to be determined by that of the metric. For
example $\Phi$ will be $C^{k+1,\alpha}$ if the metric is $C^{k,\alpha}$; smooth
or analytic if the metric is, etc. This is proved by a bootstrap argument
applied to \eq{LCBG.1} below. See~\cite[Appendix~A]{ChAscona} for the analytic
case.

\item
The maximal manifolds $(\tmcM ,\tilde \sptmetric)$ need not be unique,
and may  depend upon $W$. A non--trivial example of $W$ dependence,
with $W=C^{k,\alpha}$,
is given by a class of Robinson--Trautman (RT) space--times studied
in~\cite{ChRTsmoothness},
 \syncx{quote the relevant section if the case arises}
which for
$k+\alpha\ge 123$ do not admit any non--trivial future extensions, while for
$k+\alpha < 118$ possess an infinite number of non--isometric vacuum RT
extensions.
\end{enumerate}
\end{Remarks}

\proof
For $\ell\ge n$ let $A_\ell$ denote the  set
of subsets of
$\R ^\ell$ which are $n$--dimensional, paracompact, connected, Hausdorff, $C^1$
manifolds, set $A_\infty=\cup_{\ell=0}^\infty
A_\ell$. By a famous theorem of Whitney \cite{Whitney1} every such  manifold can be embedded in $\R
^\ell$ for some $\ell$, which shows that every manifold has a
representative which is an element of $A_\infty$. It follows that
without loss of generality a manifold
can be viewed as an element of $A_\infty$, and we shall do so.
With this definition the collection of all $C^1$ manifolds is $A_\infty$,
and therefore is a set. It follows from the axioms of set theory
that the collection of all $C^1$
manifolds which are $W$-manifolds forms a set.
Now, a Lorentzian
manifold can be identified with a subset of the bundle $T^{0,2}\mcM$ of 2--covariant tensors on $\mcM$. Next, a map $i$ from $\hyp$
to $\mcM$  can be identified
with a subset of the product $\hyp\times \mcM$. One easily concludes, using the axioms presented in e.g.~\cite[Appendix]{KelleyTopologyReprinted},
 \syncx{consider adding}
that the
collection, say $\mw$, of  Lorentzian $W$-manifolds with a  $W$-compatible embedding
of $\hyp$  forms a set.

Let
$(\mcM,\sptmetric)$ be a
Lorentzian  $W$-manifold with $W$-compatible   embedding $i:\hyp\to \mcM$. Consider
the
subset $\mwm$ of $\mw$
defined as the set of those Lorentzian manifolds $(\tmcM ,\tilde
\sptmetric,\tilde\imath)$ with   embedding of $\hyp$
for which there exists an isometric $C^1$
embedding $\Phi:\mcM\rightarrow\tmcM $ with a $W$-compatible
embedding $\tilde\imath = \Phi\circ i$.
On $\mwm$ we can
define a relation $\prec$ as follows: $\tmg\prec\tmgo$ if there
exists an isometric $C^1$ embedding $\tilde\Phi:\tmcM \rightarrow \tmcM_1$ satisfying $\tilde\Phi\circ\tilde\imath=\tilde\imath_1$.
 We claim that $\prec$ is a partial order. Here the only non-obvious property is
antisymmetry, namely if $\tmg\prec\tmgo$  and $\tmgo\prec\tmg$, then
$\tmg=\tmgo$.
So let $\tilde\Phi:\tmcM \rightarrow \tmcM_1$ and $\tilde\Phi_1:\tmcM _1 \rightarrow \tmcM$ be the relevant embeddings:
$$
\xymatrix{ && \tmcM    \ar@/^1pc/[dd]|{\tilde \Phi}  \\
\hyp \ar[r]^i  \ar@/^/[urr]^{\tilde\imath} \ar@/_/[drr]^{\tilde\imath_1}
& \mcM \ar[ur]^{  \Phi} \ar[dr]^{  \Phi_1} &\\
&&  \ar@/^1pc/[uu]|{\tilde \Phi_1} \tmcM _1 \;.}
$$
Then $\tilde\Phi_1\circ\tilde\Phi $ is an isometry of $(\tmcM ,\tilde\sptmetric) $ which is the identity on $\tilde\imath (\hyp)$.  By \eq{17IX11.1} the map $\tilde\Phi_1\circ\tilde\Phi $ is the identity on $\tmcM $, thus $\tilde\Phi\circ\tilde\Phi_1$  is the identity on $\tmcM _1$ as well, proving that $\tmg=\tmgo$ up to isometry, as desired.

If $A\subset\mwm$ is a chain, define
\bel{12IX12.1}
  \omcM  =
\left(\cup_{\tmg \in A}\tmcM \right)/\sim
 \;,
\ee
where for $p\in \tmcM$ and $q\in\tmcM_1$ we set $p\sim q$ iff $q=\Phi(p)$, where $\Phi:\tmcM \rightarrow\tmcM_1$ is the isometric $C^1$ embedding such that $\Phi\circ\tilde\imath=\tilde\imath_1$. It is not too difficult to show that
$\omcM $ is a  manifold (Hausdorff, paracompact, connected),
 and a
Lorentzian metric $\overline \sptmetric$ can be defined on $\omcM $ in the
obvious way. By the chain-inheriting property, $\omcM$ is a $W$-manifold. Since every $\tmcM $ such that $\tmg\in A$ can be embedded in
$\omcM $ as
$$
 \tmcM \ni p \mapsto [p]_\sim\in \omcM
 \;,
$$
it follows that
 $\omcM $ is an upper bound for $A$. The Kuratowski-Zorn Lemma ({\em cf. e.g.\ }\cite{KelleyTopologyReprinted}) shows that
$\mwm$  has
maximal elements, which had to be established.\hfill $\Box$

\bigskip

Before continuing, it appears useful to exhibit classes of space-times in which
condition \eq{17IX11.1} is satisfied.
The simplest case is that of $C^{k,\alpha}$ manifolds, where $k+\alpha\ge 3$, with
$C^{k-1,\alpha}$ metrics, and with $C^{1}$ submanifolds and embeddings:

\begin{Proposition}
\label{PCBG}
Let $(\mcM,\sptmetric )$ be a $C^{2,1}$,  connected Lorentzian
manifold with a $C^{1,1}$ metric, let $\Psi:\mcM\rightarrow \mcM$ be a $C^1$ map such
that
\[
\Psi^*\sptmetric  = \sptmetric \;, \qquad \Psi\big|_S= \mbox{id}  \quad (S\ne
\emptyset)\;,
\]
where $S$ is either
\begin{enumerate}
 \item
an open set, or
\item $S=\{p\}$ is a point $p\in \mcM$, in which case we further assume that
$\Psi^*(p)$ is the identity, or
\item a
$C^{1}$ submanifold of codimension $1$, in which case we moreover assume that
$\Psi$ preserves  time-orientation.
\end{enumerate}
Then
\[
\Psi=\mbox{id}\;.
\]
\end{Proposition}

\begin{Remark}
Note that each of the conditions is necessary, and that in point 1 and 3 neither
size nor completeness requirements are imposed on $S$.
\end{Remark}

\proof  Suppose first that $S$ is an open set, let $\tilde S$ be
the largest open set such that $\Psi\big|_{\tilde S}=\mbox{id}$.
Suppose that $\tilde S$ is not closed, thus there exists $p\in
\partial\tilde S$, let $\cohere$ be any neighbourhood of $p$ with a
local coordinate system such that $x^\mu (p)=0$. Continuous differentiability of
$\Psi$
implies, in local coordinates,
\be
 \label{LCBG.0}
 \Psi^\mu(0)=0\;, \quad \frac{\partial \Psi^\mu}{\partial
 x^\nu}(0)=\delta^\mu_\nu
 \;.
\ee
From $\Psi^*\sptmetric =\sptmetric $ one has
\be
\sptmetric _{\alpha\beta}(x)=\sptmetric _{\mu\nu}\left(\Psi(x)\right)
\frac{\partial \Psi^\mu}{\partial
x^\alpha}\frac{\partial \Psi^\nu}{\partial
x^\beta}\;,
\label{LCBG.2}
\ee
\be
\label{LCBG.1}
\frac{\partial^2 \Psi^\mu}{\partial
x^\alpha\partial
x^\beta}=\Gamma^\sigma_{\alpha\beta}(x)\frac{\partial \Psi^\mu}{\partial
x^\sigma}-\Gamma^\mu_{\nu\rho}\left(\Psi(x)\right)
\frac{\partial \Psi^\nu}{\partial
x^\alpha}\frac{\partial \Psi^\rho}{\partial
x^\beta}\;,
\ee
where $\Gamma$ denotes the Christoffel symbols of the metric $\sptmetric $.
Indeed, recall that \eq{LCBG.1} is obtained by differentiating \eq{LCBG.2} and
algebraic manipulations when $\Psi$ is $C^2$.

When $\Psi$ is assumed to be $C^1$
only, the same manipulations show that \eq{LCBG.1} holds in a distributional
sense. But since the right-hand side is continuous, we conclude that $\Psi$ is
$C^2$ in any case.

Setting $A^\alpha{}_\beta\equiv\frac{\partial \Psi^\alpha}{\partial
x^\beta}$, from (\ref{LCBG.1}) one obtains the following system of
ODE's along rays emanating from the origin:
\[
\frac{d\Psi^\mu}{dr} = A^\mu{}_\beta \frac{x^\beta}{r}\;, \qquad
r=\left(\sum (x^\alpha )^2\right)^{1/2}\;,
\]
\[
\frac{dA^\mu{}_\beta}{dr} =
\left(\Gamma^\sigma_{\alpha\beta}(x)A^\mu{}_\sigma
-\Gamma^\mu_{\nu\rho}\left(\Psi(x)\right)
A^\nu{}_\alpha A^\rho{}_\beta \right) \frac{x^\alpha}{r}\;.
\]
The initial conditions (\ref{LCBG.0}) together with uniqueness of solutions
of systems of ODE's imply $\Psi^\mu=x^\mu$ in $\cohere$, which leads to
a contradiction, and shows that $\partial \tilde S=\emptyset$, thus $\tilde
S=\mcM$. This proves point 1.

Note that we have also shown that if $\Psi(p)=p$ and $\Psi^*(p)=\id$, then
$\Psi =\id$ on a neighborhood $\cohere$  of $p$, hence $\Psi=\id$ by point 1, and
point 2 is proved as well.

Suppose now that $S$ is a hypersurface, let $p\in S$.  Then $\Psi^*$ is the
identity on $T_pS$ and preserves  time-orientation. Elementary
algebra shows that $\Psi^*(p)$ is the identity: Indeed, this is straightforward
if $T_pS$ is spacelike or timelike. If $S$ is null, let $n,\ell,e_A$,
$A=2,\ldots, n$, be a basis of $T_p\mcM$ such that $n$ and $\ell$ are null, the
$e_A$'s are ON and orthogonal to $\ell$ and $n$, with $\ell$ and $e_A$ tangent
to $S$. Then $\Psi^*$ is a Lorentz transformation that leaves invariant
$\ell$, the $e_A$'s,   the space spanned by $\ell$ and $n$, and preserves time-orientation, hence
is the identity. The
result follows now by point 2.
\hfill $\Box$

\medskip

Recall that\emph{wave coordinates}, often also called \emph{harmonic coordinates}, are defined by the requirement that $\Box_g x^\mu=0$,
where $\Box_g$ denotes the d'Alembert operator of a metric $g$.
We have the following  ``Lipschitz-harmonic" version of Proposition~\ref{PCBG}:

\begin{Proposition}
\label{PCBG2n}
Let $(\mcM,\sptmetric )$ be a globally hyperbolic connected Lorentzian
$(n+1)$-dimensional manifold with differentiable spacelike Cauchy
surface $\hyp$. Let $\Psi:\mcM\rightarrow \mcM$ be a time-orientation preserving $C^1$ map such
that
\[
\Psi^*\sptmetric  = \sptmetric \;, \qquad \Psi\big|_\hyp= \mbox{id} \;.
\]
If $\mcM$ can be covered by wave-coordinates patches in which
the metric is $C^{0,1}$, then
\[
\Psi=\mbox{id}\;.
\]
\end{Proposition}

\begin{Remark}
 \label{R2XI11.1}
We have chosen the wave-coordinates condition for definitess. The argument applies to any systems of coordinates in which $\Box_g x^\mu = F^\mu(x, g)$ with Lipschitz functions $F^\mu$.
\end{Remark}

\proof
\Eq{LCBG.1}, understood distributionally, in coordinates where $g$ is Lipschitz, shows that $\Psi$ is $C^{1,1}$.

Let $p\in \hyp$, since $\Psi_*$ is an isometry and leaves $T_p\hyp$ invariant, it preserves $(T_p\hyp)^\perp$. As $\Psi$ preserves time-orientation, $\Psi_*$ maps the unit normal to $\hyp$ to itself. It follows that $\Psi_*$ is the identity at $p$. Since $p$ is an arbitrary point on $\hyp$, we find that $\Psi_*$ is the identity on $\hyp$; in local coordinates, $\partial \Psi^\mu/\partial x^\nu|_\hyp = \delta^\mu_\nu$.

Let  $\mcO$ denote a conditionally compact domain of definition of some wave-coordinates in which the metric is locally Lipschitz, thus
\bel{2XI11.1}
0= \Box_g x^\mu = -g^{\alpha\beta}\Gamma^\mu _{\alpha\beta}
 \;.
\ee
Let $\hypmco \tau\subset \mcO$ denote the level set within $\mcO$ of a differentiable time function $t$:
$$
 \hypmco \tau:=\{t=\tau\}\cap \mcO
 \;.
$$
Note that we are \emph{not} assuming that $t=x^0$.

Consider a point $x$ with coordinates $x^\mu$ such that $\Psi(x)\in \mcO$.
Contracting \eq{LCBG.1} with the inverse metric, and using the wave-coordinates condition,
one obtains
\bean
 \Box_g \Psi^\mu(x)  &= & g^{\alpha\beta} (x) \left(
\frac{\partial^2 \Psi^\mu}{\partial
x^\alpha\partial
x^\beta} -\Gamma^\sigma_{\alpha\beta}(x)\frac{\partial \Psi^\mu}{\partial
x^\sigma}
 \right)
\\
 &= & - g^{\alpha\beta} (x) \Gamma^\mu_{\nu\rho}\left(\Psi(x)\right)
\frac{\partial \Psi^\nu}{\partial
x^\alpha}\frac{\partial \Psi^\rho}{\partial
x^\beta}
 \;.
\eeal{27X11.1}
Setting $\psi^\mu:= \Psi^\mu - x^\mu$,
this can be rewritten in the form
\bean
 \Box_g \psi^\mu(x)
 &= & - g^{\alpha\beta} (x) \Gamma^\mu_{\nu\rho}\left(x+\psi(x)\right)
 \left(\frac{\partial \psi^\nu}{\partial
x^\alpha}+\delta^\nu_\alpha\right)
 \left(\frac{\partial \psi^\rho}{\partial
    x^\beta}+\delta^\rho_ \beta\right)
\\
 &= & - g^{\alpha\beta} (x) \Gamma^\mu_{\nu\rho}\left(x+\psi(x)\right)
  \frac{\partial \psi^\nu}{\partial
x^\alpha}
 \left(\frac{\partial \psi^\rho}{\partial
    x^\beta}+2\delta^\rho_ \beta\right)
 \nonumber
\\
 &&
  - \big( g^{\alpha\beta} (x)
  -  \underbrace{ g^{\alpha\beta} \left(x+\psi(x)\right)
   \big)
    \Gamma^\mu_{\alpha\beta}\left(x+\psi(x)\right)}_0
 \;.
\eeal{27X11.2}
Here we have added the last, vanishing term to make it clear that the last line can be estimated, almost everywhere, by a multiple of $ |\psi|$ when the metric is Lipschitz.

To continue, we will need:
\begin{Lemma}
 \label{L28X11.1}
 If $\psi^\mu=0$ on $\hypmco \tau$, then $\psi^\mu=0$ on the domain of dependence $\mcD_J(\hypmco \tau, \mcO)$.
\end{Lemma}
\proof
The argument proceeds via a standard energy inequality, but some care is needed to take into account the low differentiability, and the fact that \eq{27X11.2} only holds in local coordinates.

Let $\hypmco {\tau,n}\subset \hypmco \tau$ be an exhaustion of $\hypmco \tau$ by compact submanifolds with smooth boundary.
Let $X$ be any differentiable timelike vector field on $\mcM$ and let $T_{\mu\nu}$ be the energy-momentum tensor associated with $\psi$,
defined as
\bel{16IX12.4}
 T_{\mu\nu}:= \sum_\alpha \big(\partial_\mu \psi^\alpha  \partial_\nu \psi^\alpha -
 \frac 12 g^{\sigma\rho} \partial_\sigma \psi^\alpha \partial_ \rho \psi^\alpha g_{\mu\nu} + \psi^\alpha \psi ^\alpha X_\mu X_\nu
  \big)
 \;.
\ee
Then $T_{\mu\nu}$ is locally Lipschitz.

Consider the domain of dependence $\mcD_J(\hypmco {\tau,n},\mcO)$. For smooth metrics, it is well known that this is a  set with Lipschitz boundary, which
is spacelike or null almost everywhere. The result remains true for continuous metrics, as seen  by noting that   $\partial \mcD_J(\hypmco {\tau,n},\mcO)$ can be written, locally, as an achronal graph. Such graphs are sandwiched between graphs of future- and past light-cones of any of their points, which easily implies the result.
 \syncx{proved as Proposition~\ref{P13IX12.1} in the extended causality notes; consider adding the argument here}

For $t\ge \tau$ define
\beaa
 &
 \Omega_{n,t}= \mcD^+_J(\hypmco {\tau,n},\mcO)\cap \{\tau\le x^0 \le t\}
\;,
 &
\\
&
 T_n = \sup_{p\in \mcD^+_J(\hypmco {\tau,n},\mcO)}t(p)
  \;,
 &
\\
&
 I_n = \{t\in[\tau,T_n]:\, \Psi=\id \ \mbox{on}\  \Omega_{n,t}\}
  \;.
 &
\eeaa
(Note that $T_n<\infty$.) The set $I_n$ is not empty, since $\tau \in I_n$. Clearly, $I_n$ is closed in $[\tau,T_n]$. We will show that $I_n$ is also open in $[\tau,T_n]$, which will prove that
\bel{27IX12.1}
 \mbox{$\psi=0$ on $\mcD^+_J(\hypmco {\tau,n},\mcO)$.
}
\ee
Let, thus $s\in I_n$. Since $\Psi$ is the identity on $\Omega_{n,s}$, there exists $T>s$ such that  $\Psi(\Omega_{n,T})\subset \mcO$. Hence \eq{27X11.2} applies on $\Omega_{n,t}$ for $\tau\le t\le T$ and so there exists a constant $C$ such that there we have, almost everywhere,
\beaa
  |\nabla_\mu T^{\mu}{}_\nu| &= &  \bigg|\sum_\alpha \Box_g \psi^\alpha \partial_\nu \psi ^\alpha+ \nabla_\mu (\psi^\alpha \psi ^\alpha X^\mu X_\nu) \bigg|
\\
   & \le &
    C( |\psi^2| + |\partial \psi|^2)
 \;.
\eeaa
As already pointed-out, the last term in \eq{27X11.2} has been estimated by $C | \psi|$ using the fact that the metric is Lipschitz-continuous. The estimation of the remaining terms is straightforward, for example  terms of the form $(\partial \psi)^3$ in $\nabla_\mu T^{\mu}{}_\nu$ have been estimated by $C |\partial \psi|^2$.

Letting
$$
 E_n(t)= \int_{\hypmco {t}\cap\, \Omega_{n,T}}T^\mu {}_\nu X^\nu dS_\nu
 \;,
$$
and using the Stokes' theorem for Lipschitz vector fields on Lipschitz domains (see Corollary~\ref{C21IX12.1}, Appendix~\ref{A21IX12.1}; compare~\cite{DriverNotes,Mattila,FedererMeasureTheory})
 one obtains, for some constant $C_n$,
$$
 \forall \ \tau\le t \le T \qquad E_n(t) \le C_n \int_\tau^t E_n(s) ds
 \;.
$$
Here we have used that $\psi=0$ on $\hypmco \tau$ and, as before, $\partial \psi=0$ on $\hypmco \tau$ as well.
Gronwall's Lemma gives $E_n(t)=0$ for $0\le t\le T$. This proves that $I_n$ is open, and establishes \eq{27IX12.1}.

The fact that $\Psi=\id$ on $\mcD^-_J(\hypmco {\tau,n},\mcO)$ is obtained in an identical way.

Since
$$
 \cup_n \mcD_J(\hypmco {\tau,n},\mcO)= \mcD_J(\hypmco {\tau},\mcO)
 \;,
$$
Lemma~\ref{L28X11.1} follows.
\qedskip

Returning to the proof of Proposition~\ref{PCBG2n}, let $h$ be any complete Riemannian metric on $\mcM$. Let $p\in \hyp$, denote by $B_p(n)$ the open $h$-distance ball centred at $p$ of radius $n$,  and let
$$
 K_n := \overline{ J^-(\overline{B_p(n)}) \cap J^+(\overline{B_p(n)})}
 \;.
$$
For smooth metrics it is a standard fact that the interior $\mathring K_n$ of $K_n$ is a globally hyperbolic  subset of $\mcM$, with Cauchy surface $\hyp\cap \mathring K_n$; this can be seen to remain true for continuous metrics using the results in~\cite{ChGrant}. It also follows from the results in~\cite{ChGrant} that  $\partial K_n$ is a Lipschitz topological hypersurface, with $\mathring K_n$ lying on one side of $\partial K_n$.
 \syncx{crossref eventually twice; the Lipschitz buisiness is proved exactly as in Proposition~\ref{P13IX12.1}, so perhaps that proposition should be restated accordingly?}
Since $B_p(n)\subset \mathring K_n$  we have
$$
 \cup_n \mathring K_n=\mcM
 \;.
$$
Let $q\in \mcM$, we want to show that $\Psi(q)=q$. There exists $n$ such that $q\in \mathring K_n$. Compactness shows that $K_n$ can be covered by a finite number of conditionally compact wave-coordinates patches $\mcU_\ell$, $\ell=1,\ldots,N(n)$, in which the metric is Lipschitz-continuous.

Choose any smooth differentiable structure on $\mcM$ compatible with the $C^1$ atlas in which the metric is continuous.
By~\cite{ChGrant} or \cite{FathiSiconolfiTime}
 \syncx{crossref eventually}
there exists a smooth Cauchy time function $t$ on $\mcM$ so that $\hyp =\{t=0\}$.
Set
\beaa
 J_n & := & \{\tau\in \R: \ \hyp_\tau \cap K_n \ne \emptyset \}
 \;,
\\
 I_n &:= & \{\tau\in J_n: \ \Psi=\mathrm{id} \ \mbox{ on } \ \hyp_\tau \cap K_n\}
 \;,
\eeaa
where $\hyp_\tau$ denotes the $\tau$-level set of $t$. Since causal curves are transverse to the $\hyp_\tau$'s, $J_n$ is a closed interval containing the origin; in fact:
$$
 J_n = [\min_{p\in K_n} t(p),\max_{p\in K_n} t(p)]
 \;.
$$
Now, $I_n\ne \emptyset $ as $0\in I_n$, and $I_n$ is clearly closed in $J_n$. We wish to show that $I_n=J_n$, hence $\Psi$ is the identity on $K_n$. For this, it remains  to show that $I_n$ is open.

%
Let then $t\in I_n$, and consider those $\mcU_\ell$'s that intersect $\hyp_t$, renumbering we can assume that this happens for $\ell=1,\ldots, N$ for some $N=N(t)$. Then $\Psi$ is the identity on  $\hyptwo{\mcU_\ell}t$ for all $\ell=1,\ldots,N$ so that, by Lemma~\ref{L28X11.1}, the map $\Psi$ is the identity on
$$
\cup_{\ell=1}^N  \mcD_J(\hyptwo{\mcU_\ell}t,\mcU_\ell )
 \;.
$$
We claim that for $t'$ close enough to $t$ we have
\bel{27IX12.2}
 \hyp_{t'}\cap K_n \subset \cup_{\ell=1}^N  \mcD_J(\hyptwo{\mcU_\ell}t,\mcU_\ell )
 \;.
\ee
Indeed, suppose that this is not the case, then there exists a sequence $t_i\to t$ and points $q_i\in \hyp_{t_i}\cap K_n$ such that $q_i$ does not belong to the set appearing at the right-hand side of the inclusion  \eq{27IX12.2}. By compactness, passing to a subsequence if necessary, $q_i$ converges to $q\in \hyp_t\cap K_n$. Then $q\in \hyptwo{\mcU_\ell}t$ for some $\ell$. But  $\mcD_J(\hyptwo{\mcU_\ell}t,\mcU_\ell )$ forms a  neighborhood of $  \hyptwo{\mcU_\ell}t$, and so $q_i\in \mcD_J(\hyptwo{\mcU_\ell}t,\mcU_\ell )$ for all $i$ large enough, which gives a contradiction.

So \eq{27IX12.2} holds for all $t'$ close enough to $t$, which establishes openness of $I_n$, and finishes the proof of
Proposition~\ref{PCBG2n}.
\qedskip

One can use~\cite[Theorem~III]{hughes:kato:marsden}
 \syncx{also: a standard symmetric hyperbolic reduction of the wave equation to a system satisfying the hypotheses of \cite[Theorem~7.2.2]{MetivierParaDiff}, obtaining thus; add more details; use an ON basis and a $\sqrt{A^0}$ symmetrizer;  this reference does NOT apply directly because he does not have the zero order term, but the zero order term can be put in the source of the energy inequality and obtain the result!}
to cover  a manifold with a $\Hssp$ metric, $s>n/2+1$, by wave-map coordinate patches.
However, when transformed to wave-coordinates, the metric will be of $\Hsmsp$-differentiability class only in general. The requirement of existence of the embedding $H^{s-1 }\subset C^{0,1}$ leads to the  threshold   $s>n/2+2$ for the applicability of Proposition~\ref{PCBG2n} for general $\Hssp$ metrics.

On the other hand, solutions of the vacuum Einstein equations can be constructed directly by patching-together domains of definition of wave-coordinates~\cite{ParlongueCRAS,PlanchonRodnianski}, and then Proposition~\ref{PCBG2n} applies without loss of differentiability for the metric when $s>n/2+1$.
 \syncx{given the current wording, one might need to make a uniqueness argument here}

\section{Global uniqueness}

\subsection{An abstract theorem}
 \label{ss14XI11.1}

In the context of $\Hssp$-Lorentzian manifolds, a hypersurface $S$ will be said to be \emph{compatible} if $S$ is a coordinate-level set of a coordinate system in which the metric is of $\Hspsp$-differentiability class.

We will prove a somewhat more general version of Theorem~\ref{T11XI11.1},
where the differentiability index $s\in\R$ is only assumed to satisfy $s>n/2$, as needed to ensure  continuity of the metric,%
\footnote{Minor variations of our conditions would be appropriate for metrics which are of $\Hssp\cap C^0$-differentiability class, with $s\le n/2$. Similarly one could use more general spaces of metrics and maps, with correspondingly modified conditions \Hone-\Athree.}
provided that the following conditions hold:

\begin{enumerate}
\item [\Hone.]
     \label{H7XI11.H}
     The harmonically-reduced vacuum Einstein equations with initial data in $H^s_\loc (S)\times H^{s-1}_\loc (S)$, $S\subset\{t=0\}\subset \R^{n+1}$ have local solutions.
     \item [\Htwohere.]
     \label{H7XI11.H2}
     Two solutions $g_1$ and $g_2$ in $\Hssp$  globally coordinatized by harmonic coordinates with the same data on $S\subset\{t=0\}\subset \R^{n+1}$ coincide on $\mcD_{J,g_1}(S)\cap \mcD_{J,g_2}(S)$.%
     \footnote{We use causal-theory terminology and notation from~\cite{ChCausalityv1}.}
      \syncx{needs proof? synchronize footnote}
\item [\Azero.]
     \label{H7XI11.0} A time-orientation-preserving $C^1$ isometry of $(\mcM,g)$, with $g\in \Hssp$, which is the identity on the initial data hypersurface is the identity everywhere.
 \item [\Aone.]
     \label{H7XI11.1}Let $\Phi$ be a $C^1$ isometry of two Lorentzian manifolds with $\Hssp$ metrics. Then $\Phi$ is of $\Hspsp$-differentiability class.
%
    \item [\Atwo.]
     \label{H7XI11.2}
     For any compatible spacelike acausal hypersurface $S$  and for any initial data
$(\psi,\chi)\in H^{s+1}_\loc(S)\times H^s_\loc(S)$
          the Cauchy problem
          $$
           \Box_g f=0\;,
           \qquad
           f|_S=\psi\;, \quad \partial f |_S = \chi
             \;,
           $$
             has a unique solution  of $\Hspsp$-differentiability class in the $J$-domain of dependence of $S$.

     \item [\Athree.]
     \label{H7XI11.3}
     If $\Psi$ is of $\Hspsp$-differentiability class and $g$ is in $\Hssp$, then $\Psi^* g$ is in $\Hssp$.
\end{enumerate}

We note that \Hone-\Htwohere\ and \Atwo-\Athree\ are needed for local existence  and uniqueness of solutions near the initial data hypersurface. Hypothesis \Azero\  is used in the proof of existence of maximal developments. The hypothesis \Aone\ is  used to extend isometries in the proof of uniqueness of maximal globally hyperbolic developments.

We claim that:

\begin{Theorem}
 \label{T7XI11.1}
 Under the remaining hypotheses of Theorem~\ref{T11XI11.1}, suppose instead that $\R\ni s>n/2$.
If moreover  the hypotheses \Hone, \Htwohere, and  \Azero -\Athree\ hold, then the conclusions of Theorem~\ref{T11XI11.1} hold.
\end{Theorem}

The key to the proof is the following proposition:

\begin{Proposition}\label{contP1}
Let $s>n/2$, let $(\mcM_a, \sptmetric_a) $, $a = 1, 2$, be vacuum globally hyperbolic
$\Hssp$ space-times with $C^1$
Cauchy surfaces $\hyp _a$ and with $\Hssp\times\Hsmsp$ initial data, and suppose that ($\mcM_2, \sptmetric_2$) is maximal.
Let $\cohere \subset \mcM_1$ be a globally hyperbolic neighborhood of
$\hyp _1$, with Cauchy surface $\hyp_1$, and
suppose there exists an injective time-orientation preserving isometry $\Psi_{\cohere} : \cohere
\rightarrow \mcM_2$,
such that $\Psi_{\cohere}(\hyp _1)$ is acausal.  If the hypotheses \Htwohere\ and \Aone-\Athree\  hold, then there exists an injective isometry
%
%
\be
\Psi : \mcM_1 \rightarrow \mcM_2,
\ee
such that $\Psi |_{\cohere} = \Psi_{\cohere}$.
\end{Proposition}

Before passing to the proof of Proposition~\ref{contP1}, let us note that
Theorem~\ref{T7XI11.1} is a corollary thereof:

\medskip

{\sc\noindent Proof of Theorem~\ref{T7XI11.1}:}
The existence of some vacuum globally hyperbolic development, and uniqueness-up-to-isometry in a globally hyperbolic neighborhood of $i(\hyp)$, are standard consequences of hypotheses
\Hone,  \Htwohere, \Atwo\ and \Athree.
 \syncx{this should be crossrefered to an appropriate section in the levoca notes}
The existence of maximal vacuum globally hyperbolic developments follows from
hypothesis \Azero\  and Theorem~\ref{TL12IX11.1}.

It should  be clear from Proposition~\ref{contP1} that two maximal vacuum globally hyperbolic developments are
isometrically diffeomorphic: Indeed, let $(\mcM_a,g_a,i_a)$, $a=1,2$, be two such developments, with $\hyp_a=i_a(\hyp)$. Let $\mcO_a\subset \mcM_a$ be isometric globally hyperbolic neighborhoods of $i(\hyp_a)$, with injective time-orientation preserving isometries $\Psi_{\mcO_1}:\mcO_1\to\mcO_2$, $\Psi_{\mcO_2}:\mcO_2\to\mcO_1$, the existence of which has just been pointed out. Proposition~\ref{contP1} provides   injective isometries
\be
\Psi_1 : \mcM_1 \rightarrow \mcM_2\;,
\quad
\Psi_2 : \mcM_2 \rightarrow \mcM_1\;.
\ee
The construction so far gives, in fact, $\Psi_a |_{\cohere_a} = \Psi_{\cohere_a}$,   $\Psi_1\circ i_1=i_2$,  and $\Psi_2\circ i_2=i_1$. This implies that $\Psi_2\circ \Psi_1$ is the identity on $\hyp_1$, and thus $\Psi_1$ and $\Psi_2$ are inverse to each other by hypothesis \Azero.
\qedskip

To complete the proof of Theorem~\ref{T7XI11.1} it remains to establish Proposition~\ref{contP1}:

\medskip

{\sc\noindent Proof of Proposition~\ref{contP1}:}
The condition that $s>n/2$ guarantees that the metric is continuous, and so the causality theory of~\cite{ChGrant} with continuous metrics applies.

Consider the collection $ {{\cal X}}$ of all pairs
$(\cuNorma ,\Psi_\cuNorma )$, where
$\cuNorma  \subset \mcM_1$ is a neighborhood
of $\hyp _1$ such that
$\hyp _1$
is a Cauchy surface for $(\cuNorma ,  \sptmetric_{1}|_{\cuNorma })$,%
\footnote{By this we mean that every future-directed future-inextendible causal curve which starts in $\cuNorma\cap J^-(\hyp_1)$ remains in $\cuNorma$ until it meets $\hyp_1$; similarly for past-directed causal curves starting in $\cuNorma\cap J^+(\hyp_1)$.}
 and where
$\Psi_{\cuNorma }:\cuNorma \rightarrow \mcM_2$ is an isometric
diffeomorphism between $\cuNorma $ and $\Psi_{\cuNorma }(\cuNorma )\subset \mcM_2$
satisfying
\bel{19IX11.1}
 \Psi_{\cuNorma }|_{\cohere} = \Psi_{\cohere}
  \;.
\ee
The collection ${\cal{X}}$ can be partially
ordered by
inclusion:
 $(\cuNorma ,
\Psi_{\cuNorma }) \leq (\cv, \Psi_{\cv})$ if  $\cuNorma  \subset \cv$ and if
$\Psi_{\cv}|_{\cuNorma } =
\Psi_{\cuNorma }$.  Let $(\cuNorma _\alpha, \Psi_\alpha)_{\alpha \in\Omega}$ be
a
chain in $ {\cal{X}}$,
set $\cw = \cup _{\alpha\in\Omega} \cuNorma _\alpha$, define $\Psi_{\cw} :
\cw \rightarrow
\mcM_2$ by  $\Psi_{\cw}|_{\cuNorma _\alpha} = \Psi_\alpha$;   clearly ($\cw,
\Psi_{\cw}$) is a
majorant for $(\cuNorma _\alpha, \Psi_\alpha)_{\alpha \in\Omega}$. As in the
proof of Theorem~\ref{TL12IX11.1}, using the
set-theory
axioms from~\cite[Appendix]{KelleyTopologyReprinted} it can be seen
that $ {{\cal X}}$ forms
a set, we can thus apply the Kuratowski-Zorn  Lemma \cite{KelleyTopologyReprinted} to infer that
there exist maximal elements in $ {{\cal X}}$.

 Let then $({\tmcM}, \Psi)$ be
any maximal element, by definition
(${\tmcM},
\sptmetric_{1}|_{{\tmcM}}$) is thus globally hyperbolic
with Cauchy
surface $\hyp _1$,
and $\Psi$ is a one-to-one isometry from ${\tmcM}$ into $\mcM_2$ such that
\be
 \Psi |_{ \cohere}  = \Psi_{\cohere}
 \;.
  \label{cont(1)}
\ee

We conclude that without loss of generality we can assume that $(\cohere,\Psi_{\cohere})$ is maximal. We then have:

\begin{Lemma}
\label{contL1}
 Under the hypotheses of Proposition \ref{contP1}, suppose that $(\cohere,
\Psi_{\cohere})$  is
maximal.  Then the topological space
$$\mcM^\prime = (\mcM_1 \sqcup \mcM_2)/\Psi_{\cohere}$$
is Hausdorff, hence a manifold.
\end{Lemma}

\begin{Remark}
  Recall that $\sqcup$ denotes the disjoint union, while $(\mcM_1\sqcup
\mcM_2)/\Psi$ is the   quotient space $(\mcM_1 \sqcup \mcM_2 )/  \sim, $ where
$p_1 \in \mcM_1$
is equivalent to $p_2 \in \mcM_2$ if $p_2 = \Psi(p_1)$, with the quotient topology.
\end{Remark}

\proof
Let $p', q' \in \mcM^\prime$ be such that there exist no open
neighborhoods
separating $p'$ and $q'$; clearly this is possible only if, interchanging $p$ with
$q$ if necessary, we have $p'=[p]$, with $p\in \partial \cohere$ and $q'=[q]$, with $q\in \partial
\Psi_{\cohere}({\cohere})$. Such points $p$, $p'$, $q$ and $q'$ will be called ``non-Hausdorff".

Let ${\cal H}$
denote the set of non-Hausdorff points $p $ in
$\mcM_1$, thus $p^\prime = i_{\mcM_1}(p)\equiv [p]$ is non-Hausdorff in $\mcM'$,
where $i_{\mcM_1}$ denotes
the    embedding of $\mcM_1$ into $\mcM^\prime$. By elementary topology ${\cal H}$ is
closed (as its complement is open),
and
we have just seen that ${\cal H} \subset  \partial \cohere $.

Suppose that  ${\cal H}   \not= \emptyset$, changing time orientation
if necessary we
may assume that ${\cal H}\cap
I^+(\hyp _1)\ne\emptyset$. Let  $\hat p  \in {\cal H} \cap
I^+(\hyp _1)$.  We wish to show
that there necessarily exists $p \in  {\cal H}$  such that
\be
 {J}^-(p) \cap  {\cal H}\cap I^+(\hyp _1) = \{ p\}. \label{cont(2)}
\ee
If (\ref{cont(2)}) holds with $p = \hat p$ we are done, otherwise consider the
(non-empty) set
$\cy$ of future directed causal paths $\Gamma: [0,1] \rightarrow I^+ (\hyp )$
such that
$\Gamma (0) \in {\cal H}, \Gamma(1) = \hat p$.  $\cy$ is directed by
inclusion:
$\Gamma_1 < \Gamma_2$ if $\Gamma_1  ([0,1])  \subset \Gamma_2 ([0,1]) $. Let
$\{\Gamma_\alpha\}_{\alpha \in \Omega}$ be a chain in ${\cal Y}$, we let
$\Gamma $ be the causal curve the image of which is the union of the images of all the
$\Gamma_\alpha ([0,1]) $'s.%
\footnote{Equivalently, we can parameterize all the $\Gamma_\alpha$'s by the negative of proper distance from $\hat p$ with respect to
some auxiliary complete Riemannian metric, $\Gamma_\alpha:[s_\alpha,0]\to \mcM$. Let
$s_*=\inf\{s_\alpha\}$, then for $s\in (s_*,0]$ we define $\Gamma(s)\in \mcM$ as $\Gamma_\alpha(s)$,
where $\Gamma_\alpha$ is any curve from the family for which $s_\alpha < s$; the result does not depend upon $\alpha$.}  Clearly $\Gamma\subset J^+(\hyp _1) =
I^+(\hyp _1) \cup  \hyp _1$,
and global hyperbolicity implies that
 \syncx{  image of a curve and curve mixed up in previous proof, corrected 27 IX 12, should be reread; elsewhere too?}
$\Gamma$ must be extendible, thus $\{\Gamma_\alpha(0)\}$ accumulates at
some $p_* \in
I^+(\hyp _1)\cup  \hyp _1$. As
 $ \cohere$ is an open neighborhood of $\hyp _1$ the case $p_*  \in
\hyp _1$ is not
possible, hence $p_*  \in I^+(\hyp _1)$ and consequently $\Gamma \in \cy$.
It follows that every chain in ${\cal Y}$ has a majorant, and by
Zorn's Lemma ${\cal Y}$
has maximal elements.  Let then $\Gamma$ be any maximal element of
${\cal Y}$, setting
$p = \Gamma(0)$ the equality (\ref{cont(2)}) must hold.

We now claim that (\ref{cont(2)}) also implies that there exists an open neighborhood $\hcV$ of $p$ such that
\be
{J}^-(p) \cap \partial \cohere \cap I^+(\hyp _1)\cap \hcV= \{ p\}. \label{cont(3)}
\ee
In order to establish \eq{cont(3)}, we start with the following lemma
(recall that our notation and terminology follow~\cite{ChGrant}):
 \syncx{synchronize comments about terminology if not
stand-alone}

\begin{Lemma}
 \label{L21IX11.1}
Let $p\in \partial\cohere\cap J^+(\hyp_1)$, then $\cI^-(p)\cap J^+(\hyp_1)\subset
\cohere$.
\end{Lemma}

\begin{remark}
For $C^{0,1}$ metrics one has $I^-(p)=\cI^-(p)$~\cite{ChGrant},
 \syncx{synchronize}
and then the result is standard. We do not know whether the inclusion $I^-(p)\cap J^+(\hyp_1)\subset
\cohere$ holds for metrics which are merely continuous.
\end{remark}

\proof
Let $q\in \cI^-(p)\cap J^+(\hyp_1)$,
then $p\in \cI^+(q)$, and so $\cI^+(q)$
forms an open neighborhood of $p$. Let $p_i\in \cohere$ be a sequence converging
to $p$, then $p_i \in \cI^+(q)$ for $i$ large enough. Let $\gamma_i$ be a
timelike curve from $p_i$ to $q$, note that $\gamma_i$ does not meet $\hyp_1$
since $q\in I^+(\hyp_1)$ and $\hyp_1$ is achronal. Let $\hat\gamma_i$ be any
past inextendible causal extension of $\gamma_i$. Global hyperbolicity implies that
$\hat\gamma_i$ is included in $\cohere$ at least until it meets $\hyp_1$ when followed
to the past from $p_i$, hence $q\in \cohere$.
\qedskip

Returning to the proof of   \eq{cont(3)},
let $\gammaXp$ be a past-inextendible locally uniformly timelike curve with
$p=\gammaXp(0)$; by global hyperbolicity and Lemma~\ref{L21IX11.1} the curve $\gammaXp $ meets $\hyp_1$ at, say, $\gammaXp(1)$. Thus $p\in \cI^+(\gammaXp(1))$; since the last set is open, there exists an open neighborhood  $\hcV$ of $p$ contained in  $ \cI^+(\gammaXp(1))$.

Suppose that \eq{cont(3)}
is wrong, then there exists a point
$$
 q\in
{J}^-(p) \cap \partial \cohere \cap I^+(\hyp _1)\cap \hcV \subset {J}^-(p) \cap \partial \cohere \cap \cI^+(\gammaXp(1))
$$
which is distinct from $p$.

Let
$\gammaXq  $ be a past inextendible $g$-causal curve starting at
$q$. By
Lemma~\ref{L21IX11.1} points on $\gammaXp $ distinct from $p$, and on $\gammaXq  $ distinct from $q$, and lying to the
future of $\hyp_1$ are
in $\cohere$.

Let $q_i\ne q$ be any sequence of points on $\gammaXq  $ converging to $q$
such that  $q_{i+1}\ni \cI^+_{g_1}(q_i)$. In particular $q_i \in\cohere$.
The
aim of the argument is to show that $\Psi_\cohere(q_i)$ has a limit in $\mcM_2$, which
will imply that $q\in\cal H$, a contradiction with \eq{cont(2)}.
The  standard
proof~\cite{ChCausalityv1,ChGrant}
 \syncx{synchronize}
of  existence of the limit of the sequence $\Psi_\cohere(q_i)$  for $C^{0,1}$
metrics uses Lemma~\ref{L21IX11.1} together with the fact that a causal curve
which is not
everywhere null can be deformed, with end points fixed, to a timelike one.
However, there exist continuous Lorentzian metrics and causal curves for which no such deformations exist~\cite{ChGrant},
 \syncx{synchronize}
and a different line of thought is needed.

By~\cite{FathiSiconolfiTime}\syncx{(compare~\citexG{Theorem~}{2.7}{T9IX11.1})}
%
%
%
there exists a smooth metric $\hg_1\succ g_1$ on $\mcM_1$
so that $\mcM_1$ is globally hyperbolic with Cauchy surface $\hyp_1$. For $i\ge 2$
let $\hg_i$
be any sequence of smooth metrics converging locally uniformly to $g_1$ such
that
$$
 \hg_1\succ \hg_{i} \succ \hg_{i+1} \succ g_1
 \;.
$$
Then all the spacetimes $(\mcM_1,\hg_i)$ are globally hyperbolic with Cauchy
surface
$\hyp_1$.

For any $i\ge 2$ and $j\in \N$, the closed null Lipschitz hypersurfaces $\partial J^+_{\hg_i}(q_j)$
separate
$\mcM_1$, with $\hyp_1$ lying to their past. Moreover, $p$ cannot lie to the $\hg_i$-causal past of $\partial J^+_{\hg_i}(q_j)$: Otherwise, there would be  a $\hg_i$-timelike curve from $\partial J^+_{\hg_i}(q_j)$ to $p$, which could be concatenated with the $g$-causal curve from $p$ to $q$ (which is $\hg_i$-timelike), and then deformed to a $\hg_i$-timelike curve from $\partial J^+_{\hg_i}(q_j)$ to $p$.  Hence either $p\in\partial J^+_{\hg_i}(q_j)$, or $p$ lies to the  timelike future of $\partial J^+_{\hg_i}(q_j)$. The former case cannot occur, since for the smooth globally hyperbolic metrics satisfying $\hg_i\succ \hg_{i+1}$ one has
$$\partial J^+_{\hg_{i+1}}(q_j)\setminus \{q_j\}\subset I^+_{\hg_{i}}\big(\partial J^+_{\hg_{i}}(q_j)\big)
 \;.
$$

The curve $
\gammaXp $ intersects each of the  $\partial J^+_{\hg_i}(q_j)$'s: indeed, $\gammaXp$ has to exit the compact set $J^-_{\hg_i}(p)\cap J^+_{\hg_i}(q_j)$; since $\partial J^-_{\hg_i}(p)$ is achronal, it can only do so through $\partial J^+_{\hg_i}(q_j)$. One can then
construct a $\hg_{i}$-causal curve $\gamma_{j,i}$ from $p$ to $q_j$ by following
$
\gammaXp $ from $p$ to its intersection point with   $\partial J^+_{\hg_i}(q_j)$,
and then following a generator of  $\partial J^+_{\hg_i}(q_j)$ until $q_j$ is
reached.
For each $j$ the curves $\gamma_{j,i}$ are $\hg_{1}$-causal, and $(\mcM_1,\hg_1)$
is globally hyperbolic, therefore there exists a $\hg_{1}$-causal curve
$\gamma_j$ from $q_j$ to $p$ which is an accumulation curve of the
$\gamma_{j,i}$'s. The curve $\gamma_j$ is $g_1$-causal
by~\citexG{Theorem~}{1.6}{T1IV11.1}.
%
%

By Lemma~\ref{L21IX11.1} the curves $\gamma_j$  are included in $\mcO$ except for their end-point $p$.
It is convenient to parameterize the $\gamma_j$'s by distance from $q_j$
with respect to an auxiliary complete Riemannian metric on $\mcM_2$. Let $s_i$ be defined as $p=\gamma_i(s_i)$.

Denote by $q\in \mcM_2$ the non-Hausdorff partner of $p$.
 Then the curve in
$\mcM_2$ defined as
$$
 \hat\gamma_i:= \Psi_\cohere\circ\gamma_i|_{[0,s_i)}
$$
is a $g_2$-causal curve lying in the compact set (see \citex{Theorem~}{2.9.9}{Tdd1})
%
%
%
$$
 J^-_{g_2}(q)\cap J^+_{g_2}\big(\Psi_\cohere(q_1)\big) \subset \mcM_2
 \;.
$$
Hence $\hat \gamma_i$ has an accumulation point, say $\hat r_i$, lying on the
boundary of $\Psi_\cohere(\cohere)$. The points $r_i$ and $\hat r_i$ form a
non-Hausdorff pair, which is only compatible with \eq{cont(2)} if $\hat r_i =
q$. So, in fact, $\hat \gamma_i$ can be extended
to a causal curve from $\Psi_\cohere(q_i)$ to $q$ by adding the end point.
We will denote by the same symbol that extension.

By global hyperbolicity of $\mcM_2$, passing to a subsequence if necessary, the
sequence $\hat\gamma_i$ accumulates at a $g_2$-causal curve $\hat\gamma$. This
shows that the sequence
$$
 \Psi_\cohere(q_i) = \hat \gamma_i(0)
$$
has a limit point in $\mcM_2$. Hence $q\in \cal H$, which contradicts
\eq{cont(2)}. We conclude that  \eq{cont(3)} must hold.

To continue, let $p_1\in \mcM_1$, $p_2\in \mcM_2$,
be
any
non-Hausdorff pair in $\mcM^\prime$ such that (\ref{cont(2)}) holds with $p =
p_1$.
We wish  to show that the isometry $\Psi_\mcO$ can be extended near $p_1$, which will lead to a contradiction. For this, around
$p_2$ we can construct
harmonic coordinates $y^\mu$   as follows:
Let $z^\mu$ be local coordinates
defined in some neighborhood $\cohere_2$ of $p_2$, such that  the metric coefficients are of $\Hssp$-differentiability class; such coordinates will be said to be \emph{$\Hssp$-compatible}. We can, and will, further assume that $z^0(p_2)=0$,
and that the level sets of $z^0$ are spacelike and acausal
near $p_2$. Set
$$
 \ci_\tau
 = \{q\in\cohere_2
 : z^0(q) = \tau\}
 \;.
$$
Passing
to a subset of $\cohere_2$ if necessary we may assume that $\cohere_2$ is
globally hyperbolic
with Cauchy surface $\ci_{0}$. By hypothesis \Atwo\  there exist functions
 $y^\mu\in \Hspsp$, (unique) solutions of the problem
%
$$
\Box_{g_{2}} y^\mu = 0,
$$
\be
y^0\big|_{\ci_{0}} = 0, \quad {\partial y^0\over\partial
z^0}\bigg|_{\ci_0} =
1, \quad
y^i\big|_{\ci_0} = z^i, \quad
{\partial y^i\over \partial z^0}\bigg|_{\ci_0} = 0
 \;.
\ee
Passing once
more to a globally hyperbolic
subset of $\cohere_2$ if necessary, the functions $y^\mu $ form a coordinate
system on $\cohere_2$.

\begin{figure}[tbh]
 \begin{center}
 \includegraphics[width=.5\textwidth]{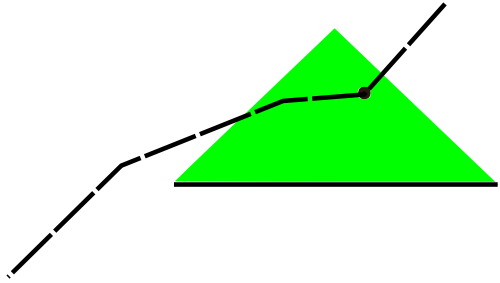}
\caption{\label{F14XI11.1} Extending the isometry $\Psi_\cohere$ near a spacelike point of $\partial\cohere$. The point $p_1$ is located at the dot, the dashed line is $\partial\cohere$, the set $\cohere$ lies under that line; the light cones have $45^\circ$ slopes. The shaded region (green in a version with colours) is the future domain of dependence of $\hat \ci$.}
\end{center}
\end{figure}

Let $w^\mu$ be any $\Hssp$-compatible coordinates near $p_1$ with domain of definition $\cV\subset \hcV$, with $\hcV$ as in \eq{cont(3)}.
We can choose $\epsilon > 0$ such that (see Figure~\ref{F14XI11.1})
\begin{enumerate}
\item $
\mcD^+_J(\ci_{
-\epsilon})
\subset
\cohere_2$,
\item $p_2 \in \mbox{int} 
\mcD^+_J(\ci_{-\epsilon})
$,
\item
$\overline{\ci_{
-\epsilon}} \subset \Psi_\cohere(\cohere) $.
\end{enumerate}

Choose  any smooth spacelike acausal hypersurface  ${\hat \ci}$ included in $\cV$ such that
$$
 \overline {\hat \ci} \subset \mcD^+_J(\Psi^{-1}_\cohere (\ci_{
-\epsilon})) \ \mbox{ and } \ p_1\in \mcD^+_J(\hat \ci)
 \;.
$$
Now, $\Psi^{-1}_\cohere$ is of $\Hspsp$-differentiability class by hypothesis \Aone.
We can thus invoke hypothesis \Atwo\ to define on  $\mcD_J({\hat \ci}) $ the functions $x^\mu \in \Hspsp(\mcD_J(\hat \ci))$ as the unique
solutions of the problem%
%
%
%
$$
\Box_{g_{1}}x^\mu = 0,
$$
$$
x^\mu\big |_{\hat \ci}  = y^\mu\circ  \Psi_\cohere\big |_{{\hat\ci} },
\quad {\partial x^\mu\over \partial
 {\hat n} }\Big|_{\hat\ci}= {\partial \left(y^\mu \circ\Psi_\cohere\right)\over \partial
\hat n
 } \Big|_{\hat\ci}
  \;,
$$
where ${\partial\over\partial {\hat n} }$ is the derivative in the direction
normal to
${\hat \ci} $.

By isometry-invariance and by the uniqueness part of hypothesis A2 we have
\be
x^\mu|_{\mcD_J({\hat\ci} )\cap \cohere} = y^\mu \circ \Psi_\cohere|_{\mcD_J(
{\hat \ci} )\cap\cohere}
 \;.
\label{cont(NEQ.1)}
\ee
Equivalently, when expressed in terms of local coordinates $x^\mu$ near $p_1$ and $y^\mu$ near $p_2$, the map $\Psi_\mcO$ is the identity on $\mcD_J(
{\hat \ci} )\cap\cohere$. In particular the $x^\mu$'s form a coordinate system on $\mcD_J(
{\hat \ci} )\cap\cohere$. Since $\Psi_\mcO$ is an isometry by hypothesis, on  $\mcD_J(
{\hat \ci} )\cap\cohere$ the metric functions for the metric $g_1$, when expressed in the  coordinates $x^\mu$, coincide with the metric functions   for the metric $g_2$, when expressed in the  coordinates $y^\mu$.

As already emphasised, the functions $x^\mu$ form a coordinate system on $\mcD_J(
{\hat \ci} )\cap\cohere$. However, we need more, namely that the $x^\mu$'s form a coordinate system near $p_1$. For this we note that on $\mcD_J({\hat\ci} )\cap \cohere$ we  have, by \eq{cont(NEQ.1)},
\beaa
  g_1{}(w) _{\mu\nu} dw^\mu dw^\nu
   &= &
     g_2{}(y(w)) _{\alpha\beta} \frac{\partial y^\alpha}{  \partial w^\mu} \frac{\partial y^\beta}{  \partial w^\nu} dw^\mu dw^\nu
\\
 &= &
     g_2{}(y(w)) _{\alpha\beta} \frac{\partial x^\alpha}{  \partial w^\mu} \frac{\partial x^\beta}{  \partial w^\nu} dw^\mu dw^\nu
 \;,
\eeaa
hence
$$
 \det \Big(\frac{\partial x^\alpha}{  \partial w^\mu} \Big)^2\Big|_{\mcD_J({\hat\ci} )\cap \cohere} =
  \frac{ \det \left(g_1(w) _{\mu\nu}\right)}{ \det \left( g_2{}(y(w)) _{\alpha\beta} \right)}\Big|_{\mcD_J({\hat\ci} )\cap \cohere}
  \;.
$$
Since the right-hand side is uniformly bounded away from zero,  continuity shows that
$\det \frac{\partial x^\alpha}{  \partial w^\mu}$ does not vanish at $p_1$.  By the implicit function theorem there exists a neighborhood
$$\calW\subset\mcD^+_J({\hat \ci} )
$$
of $p_1$ such that the map
$ \calW\ni w^\mu \mapsto x^\mu$ is a diffeomorphism onto its image.

Let
$$
 \wci_t:=\{x^0=t\}\subset \calW
$$
Making $\calW$ smaller if necessary, we can choose $\eta>0$ small enough so that
\begin{figure}[tbh]
 \begin{center}
 \includegraphics[width=.5\textwidth]{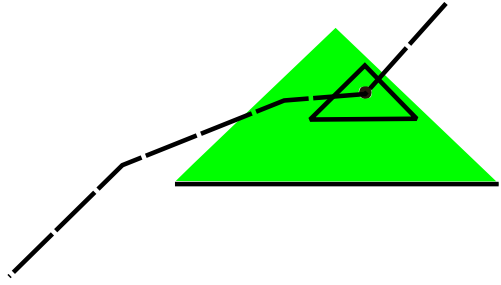}
\caption{\label{F14XI11.2} The functions $x^\mu$ form a wave-coordinates system on the future domain of dependence
of $\wci_{-\eta}$, which is represented by the triangle around the dot representing $p_1$.}
\end{center}
\end{figure}
$
 \wci_{-\eta} $
satisfies (see Figure~\ref{F14XI11.2})
\begin{enumerate}
\item $p_1 \in \mbox{int} 
\mcD^+_J(\wci_{-\eta})
$,
\item $\overline{\wci_{
-\eta}} \subset \cohere
$.
\end{enumerate}

Set
$$
 \cuNorma  = \cohere \cup
\mcD^+_J(\wci_{-\eta})
 \;,
$$
and for $p\in\cuNorma $ define
\bel{7XI11.1}
\Psi_\cuNorma (p) = \cases{\Psi_\cohere(p), & $p\in\cohere$,\cr
  q: \mbox{where $q$ is such that\ } x^\mu(q) = y^\mu (p), &$p\in
\mcD^+_J(\wci_{-\eta})\setminus \cohere$.\cr}
\ee
{}From what has been said, \eq{7XI11.1} defines a $\Hspsp$ map from
$\cuNorma $ to $\mcM_2$.

Clearly $\cuNorma $ is a globally hyperbolic neighborhood of
$\hyp _1$,
 \syncx{maybe more details useful here?}
and $\hyp _1$
is a Cauchy surface for $\cuNorma $.  Note that $\cohere$ is a proper subset of
\/$\cuNorma $, as
$p_1\in
\mcD^+_J(\wci_{-\eta}) $ but $p_1 \not\in \cohere$.

By construction
the metric $\Psi_\cuNorma^* g_2$ is of  $\Hssp$-differentiability class. (This holds by hypothesis
on $\cohere$, since there $\Psi_\cuNorma^* g_2$ coincides with $g_1$. This holds away from $\cohere$ as well, since there the map $\Psi$ is the identity in local coordinates, and in those the metric has already been shown to be in $\Hssp$).  Near $\wci_{-\eta}$ we have $\Psi_\cuNorma^* g_2=\Psi_ * g_2=g_1$ , thus $\Psi_\cuNorma^* g_2$ coincides with $g_1$ there. It follows from  hypothesis \Htwohere\ that $\Psi_\cuNorma^* g_2$  coincides with $g_1$ on $\cuNorma\cap \mcD_J^+(\wci_{-\eta})$.  So $\Psi_\cuNorma$ is a local isometry.

To prove
that $\Psi_\cuNorma $ is one-to-one, we proceed by contradiction, and consider $p, q \in \cuNorma $, $p\ne q$, such that
\bel{19IX11.6}
 \Psi_\cuNorma (p) =
\Psi_\cuNorma (q)
 \;.
\ee
Since $\Psi_{\cohere}$ is one-to-one, and since the map
\bel{19IX11.7}
 x^\mu\mapsto y^\mu
\ee
constructed above in local coordinates is one-to-one, \eq{19IX11.6} can only
occur with $p\ne q$ if $p  $ lies in the domain of the map \eq{19IX11.7} and $q$
lies in  $\cohere$, or vice-versa.
Exchanging $p$ and $q$ if necessary, we
only need to consider the former case, and note that $p\not\in\cohere $ since then $\Psi_\cuNorma$
would coincide with $\Psi_{\cohere}$ near $p$, and would therefore be injective
there. So $p$ must lie in the complement of $\cohere$, but $\Psi_\cuNorma(p)$ must
lie in $\Psi_{\cohere}(\cohere)$.

Consider a past directed timelike curve $\Gamma_1$ entirely contained in $\cohere$,
inextendible in $\cohere$, and passing through $q$. Set
$\Gamma:=\Psi_\cohere(\Gamma_1)$. Since the map
\eq{19IX11.7} is a local diffeomorphism, we can invert it locally to obtain a
pre-image of $\Gamma$ which is a past-directed timelike curve $\Gamma_2$ through
$p$. Suppose that $\Gamma_2$   meets $\ci_{-\epsilon}\subset \cohere$
when followed to the past.
Since $\Psi_\cohere$ is one-to-one, the
part of $\Gamma_2$ that lies in $\cohere$ must coincide with $\Gamma_1$, which is
not possible since $\Gamma_1$ has an end-point at $q$, while $\Gamma_2$ leaves
$\cohere$ through $\partial\cohere$. We infer that $\Gamma_2$ stops before  meeting
$\ci_{-\epsilon}\subset \cohere$ when
followed to the past. So global hyperbolicity implies that $\Gamma$ must meet
$\hyp_1$ when followed to the future. One can then construct a timelike curve
from $\Psi_\cohere(\hyp_1)$ to itself by following the image by $\Psi_\cuNorma$ of
any causal curve from $\hyp_1$ to $\Psi_\cohere(p)$, and
then $\Gamma$ from $\Psi_\cohere(p)$ to $\Psi_\cuNorma$, which is not possible as we assumed that $\Psi_\cuNorma$ is achronal.
This shows that no distinct points $p$ and $q$ satisfying \eq{19IX11.6}
exist, and we conclude that $\Psi_\cohere$ is injective, as desired.

We have thus shown, that $(\cohere, \Psi_\cohere) \leq (\cuNorma , \Psi_\cuNorma )$
and $(\cohere, \Psi_\cohere) \neq (\cuNorma , \Psi_\cuNorma )$ which contradicts
maximality  of $(\cohere, \Psi_\cohere)$.  It follows that
$\mcM^\prime$ is Hausdorff, establishing Lemma~\ref{contL1}.   $\hfill\Box$

\bigskip

Returning to the proof of Proposition~\ref{contP1},  let (${\tmcM},\Psi$)
be maximal. If
${\tmcM}=\mcM_1$ we are done, suppose then that ${\tmcM}\neq \mcM_1$. Consider the
topological space
$$
\mcM'=(\mcM_1\sqcup \mcM_2)/\Psi \;.
$$
Then $\mcM'$ is Hausdorff by Lemma \ref{contL1}, hence a manifold.

We claim that $\mcM'$ is globally
hyperbolic
with Cauchy surface
$$
 \veps'=i_{\mcM_2}(\veps_2)\approx \veps_2
$$
(recall that
$i_{\mcM_a}$ denotes the  canonical embedding of $\mcM_a$ in $\mcM'$).
Indeed, let $\Gamma'\subset \mcM'$ be an inextendible causal curve in $\mcM'$, set
$$
 \Gamma_1=i_{\mcM_1}^{-1}(\Gamma'\cap i_{\mcM_1}(\mcM_1))
 \;,
 \quad
 \Gamma_2=i_{\mcM_2}^{-1}(\Gamma'\cap i_{\mcM_2}(\mcM_2))
 \;.
$$
Clearly
$\Gamma_1\cup \Gamma_2\neq\phie$, so that either $\Gamma_1\neq\phie$, or
$\Gamma_2\neq\phie$, or both. Let  the index $a$ be such that
$\Gamma_a\neq\phie$.
If $\hat\Gamma_a$ were an extension of $\Gamma_a$ in $\mcM_a$, then
$i_{\mcM_a}(\hat\Gamma_a)$ would be an extension of $\Gamma'$ in $\mcM'$, which
contradicts maximality of $\Gamma'$, thus $\Gamma_a$ is inextendible. Suppose
that $\Gamma_1\neq\phie$; as $\Gamma_1$ is inextendible in $\mcM_1$ we must have
$\Gamma_1\cap\veps_1=\{p_1\}$ for some $p_1\in \veps_1$.
We then have
$\Psi(p_1)\in \Gamma_2$, so that
it always holds that $\Gamma_2\neq\phie$.
By global hyperbolicity of $ \mcM_2$ and inextendibility of $\Gamma_2$ it follows
that $\Gamma_2\cap\veps_2=\{p_2\}$ for some $p_2\in \veps_2$, hence
$\Gamma'\cap i_{\mcM_2}(\veps_2)=\{i_{\mcM_2}(p_2)\}$. This shows that
$i_{\mcM_2}(\veps_2)$
is a Cauchy surface for $\mcM'$, thus $\mcM'$ is globally hyperbolic.

As $\tmcM\neq \mcM_1$ we find that  $ \mcM_2$ is a proper subset of $\mcM'$, which contradicts maximality of $\mcM_2$. It
follows that we must have ${\tmcM}=\mcM_1$, and Proposition~\ref{contP1} follows.
\hfill$\Box$

\subsection{Proof of Theorem~\protect\ref{T11XI11.1}}

We are ready now to pass to the proof of Theorem~\ref{T11XI11.1}, which will occupy the remainder of this section.
In view of Theorem~\ref{T7XI11.1}, we need to check that conditions \Hone--\Athree, p.~\pageref{H7XI11.H}, are satisfied when $\N\ni s>n/2+1$.

The hypothesis \Hone\ holds by~\cite{hughes:kato:marsden}, while \Htwohere\ can be established using energy arguments along the lines of the proof of Lemma~\ref{L28X11.1}.

The hypothesis \Azero\  follows from the embedding $\Hssp\subset C^{0,1}$ for $s>n/2+1$.

\syncx{the file tryagain.tex has the space proof, and Hsspacetime has the attempt to a spacetime proof; there one can also find the simplified Parlongue argument in dimension 3}

Condition \Aone\ will be the contents of  Proposition~\ref{PL5XI11.1} where, for definiteness, the reader might want to choose the functions $F^\mu$ appearing in \eq{6IV13.3} to be zero:

\begin{Proposition}
 \label{PL5XI11.1}
Consider a local diffeomorphism $\Psi$ of $C^1$-differentiability class, and a vacuum metric $g$  of $\Hssp$-differentiability class in a coordinate system $y^\mu$, with $\N\ni s>n/2+1$. Let
$$
 \overline g:= \Psi^* g
 \;.
$$
If
\bel{6IV13.3}
 \Box_g y^\mu = F^\mu(g,y)
 \;,
\ee
for some functions $F^\mu$  smooth in their arguments,
and if
$\overline g$ is also of $\Hssp$-differentiability class with respect to
a coordinate system $x^\mu$, then
\bel{5XI11.1}
 \Psi \in  \Hspsp
 \;.
\ee
\end{Proposition}

\proof
%
%
Let $ \hyp_\tau$ denote the level sets of $x^0$ and let $\ol\hyp_\tau$ denote the level sets of $y^0$. We start by noting that for all multi-indices $0 \le |\alpha|\le s$ we have
\bel{5IV13.101}
 \partial_x^\alpha g_{\mu\nu} \in L^2_\loc(\ol\hyp_\tau)
 \;,
\ee
as needed to be able to use Proposition~\ref{sploclcompCl} of Appendix~\ref{splocAppAGC}.
This is clear for $0\le |\alpha|\le 1$ since the metric is $C^1$ by hypothesis. For higher derivatives, this follows from the usual higher energy estimates for the wave equation satisfied by vacuum metrics when  \eq{6IV13.3} holds (cf., e.g., \cite{YCB:GRbook}), by integrating the current vector field over globally hyperbolic domains, the future boundaries of which form a covering of $\ol \hyp_\tau$;  see Remark~\ref{R6IV13.1} of Appendix~\ref{A21IX12.1}.\syncx{say more here}

Let $\mcU_x$ denote the domain of definition of the coordinates $x^\mu$.
In local coordinates so that $y^\mu = \Psi^\mu(x^\alpha)$, \eq{LCBG.2}-\eq{LCBG.1} take the form
\be
\overline g _{\alpha\beta}(x)=\sptmetric _{\mu\nu}\left(\Psi(x)\right)
\frac{\partial \Psi^\mu}{\partial
x^\alpha}\frac{\partial \Psi^\nu}{\partial
x^\beta}\;,
\label{LCBG.2xc}
\ee
\be
\label{LCBG.1xc}
\frac{\partial^2 \Psi^\mu}{\partial
x^\alpha\partial
x^\beta}=\overline\Gamma^\sigma_{\alpha\beta}(x)\frac{\partial \Psi^\mu}{\partial
x^\sigma}-\Gamma^\mu_{\nu\rho}\left(\Psi(x)\right)
\frac{\partial \Psi^\nu}{\partial
x^\alpha}\frac{\partial \Psi^\rho}{\partial
x^\beta}\;,
\ee
where the $\overline\Gamma^\sigma_{\alpha\beta}$'s are the Christoffel symbols of $\overline g$.
Since both $g$ and $\overline g$ are in $C^{0,1}$, as in the proof of Proposition~\ref{PCBG2n} we have $\Psi\in C^{1,1}(\mcU_x)=W^{2,\infty}(\mcU_x)$.

For $u<s-1-n/2$
we have $H^{s-1}_\loc \subset C^u$ in dimension $n$, and a
straightforward bootstrap of \eq{LCBG.1xc} shows that
$$
  \Psi \in C^{j_m}(\mcU_x)\subset W^{j_m,\infty}_\loc(\mcU_x)=  W^{j_m,\infty}_\spaceloc(\mcU_x)\;,
$$
where $j_m$ is the largest integer strictly less than $s+1-\frac n2 $.
Let $u_m$ be the largest integer strictly less than $s-1-\frac n2$, thus $\Psi\in C^{u_m+2}$.

Suppose, first, that $u_m+1\ne s-1-\frac n2$. For $1\le j \le s-1-u_m$ we then have
\bel{8XI11.11}
 H^{s-1}_\loc \subset \cap _{1\le j \le s-1-u_m} W^{u_m+j,v_j}_\loc
 \;,
 \
 \mbox{where $v_j = \frac{2n}{n-2(s-1-u_m-j)}>2$}
 \;.
\ee
Suppose  that
\bel{8XI11.12}
 \Psi\in \Wujvjmsp   \ \mbox{for some   $j\in \N$ satisfying $2\le j   \le s -u_m $.}
\ee
Since $C^{u_m+2}\subset\Wujvjmsp$, we have shown that \eq{8XI11.12} is true for $j=2$. One verifies
that for $s>1+n/2$
Lemma~\ref{sploclcompCl} applies with 
$$
 \mbox{$(\ell,q)=(u_m+j,v_j)$  and $(k+1,p)=(u_m+j,v_{j-1})$,}
$$
establishing that   the map
$x\mapsto \Gamma^\mu_{\nu\rho}\left(\Psi(x)\right) $
is in $\Wujvjsp$.
We can thus apply Lemma~\ref{sploclproductCl} with $p=v_{j-1}$, $q=v_j$, $k=\ell=u_m+j$,   and $m=1$ to conclude that the map
$$
 x\mapsto \Gamma^\mu_{\nu\rho}\left(\Psi(x)\right)
\frac{\partial \Psi^\nu}{\partial
x^\alpha} 
$$
is in $\Wujmvjsp$. One similarly finds that the maps
$$
\mbox{$x\mapsto \Gamma^\mu_{\nu\rho}\left(\Psi(x)\right)
\frac{\partial \Psi^\nu}{\partial
x^\alpha}\frac{\partial \Psi^\rho}{\partial
x^\beta}$  and $x\mapsto \overline\Gamma^\sigma_{\alpha\beta}(x)\frac{\partial \Psi^\mu}{\partial
x^\sigma}$}
$$
 are in $\Wujmvjsp$. It follows from \eq{LCBG.1xc} that $\Psi \in \Wujpvjsp
$.
In a finite number of steps one obtains~\eq{5XI11.1}. In particular the proof is complete for odd space-dimensions $n$.

For even $n$ we necessarily have $s\ge 2+n/2$, and it remains to consider the case $u_m+1= s-1-\frac n2$. For $1\le j \le s-1-u_m$ we then have
$$
 H^{s-1}_\loc \subset
 \left\{
   \begin{array}{ll}
    \cap_{p\in [1,\infty)} W^{u_m+1,p}_\loc\;, & \hbox{$ u_m=0 $;} \\
 \cap _{2\le j \le s-1-u_m} W^{u_m+j,v_j}_\loc\cap_{p\in [1,\infty)} W^{u_m+1,p}_\loc \;, &
\hbox{$u_m\ge 1 $,}
   \end{array}
 \right.
$$
with $v_j$ as in \eq{8XI11.11}. Recall that we already know that 
$$
 \Psi \in C^{u_m+2}\subset W^{u_{m}{+2},p}_\myspaceloc
$$
for any $p\in \R$.  We can thus choose $p$ large enough so that Lemma~\ref{sploclcompCl}  with $(\ell,q)=(u_m+2,v_2)$, and $(k+1,p)=(u_m+2,p)$ applies, establishing that
the map
$$
 x\mapsto \Gamma^\mu_{\nu\rho}\left(\Psi(x)\right) 
$$
is in $\Wujvjsp$ with $j=2$.
By Lemma~\ref{sploclproductCl} with $p=q=v_j$, $k=\ell=u_m+j$,  $j=2$ and $m=1$ the map
$$
 x\mapsto \Gamma^\mu_{\nu\rho}\left(\Psi(x)\right)
 \frac{\partial \Psi^\nu}{\partial
 x^\alpha} 
$$
is in $\Wujmvjsp$. One similarly finds that the maps 
$$
 \mbox{$x\mapsto \overline\Gamma^\sigma_{\alpha\beta}(x)\frac{\partial \Psi^\mu}{\partial
x^\sigma}$ and
$x\mapsto \Gamma^\mu_{\nu\rho}\left(\Psi(x)\right)
\frac{\partial \Psi^\nu}{\partial
x^\alpha}\frac{\partial \Psi^\rho}{\partial
x^\beta}$}
$$
are in $\Wujmvjsp$ with $j=2$. It follows from \eq{LCBG.1xc} that 
$$\Psi \in \Wujpvjsp\subset
\Wujpvjpsp
$$
with $j=2$. One can now continue the previous induction argument to obtain~\eq{5XI11.1}.
\qedskip

The hypothesis \Athree\ follows from  Proposition~\ref{splocpinvnew}, Appendix~\ref{splocAppAGC}.

To verify A2, near $S$ we transform the metric to a $\Hspsp$ coordinate system  where $S$ is given by the equation $\{x^0=0\}$. The metric is of $\Hssp$-differentiability class in this coordinate system by \Athree, which has already been established. We can then obtain a local solution in local coordinates by \cite{hughes:kato:marsden}. In the overlap the solutions coincide by an energy argument, as in the proof of Lemma~\ref{L28X11.1}. The globalization to the whole domain of dependence, within a single coordinate chart, is then standard.
 \syncx{give proof?}

This completes the proof of Theorem~\ref{T11XI11.1}.
\qed

\syncx{
\input{../JamesGrant/MaximalityCriteriaFromSCC}

\input{../JamesGrant/counterexamples}

\input{../JamesGrant/Isometries}
}

\appendix

\syncx{file energy.tex}
\section{Energy inequalities and  Stokes theorem}
 \label{A21IX12.1}

In the main body of the paper we need a version of the Stokes theorem where the objects involved are poorly differentiable:%
\footnote{I am grateful to David Parlongue for useful discussions and bibliographical advice concerning this section.}

\begin{Proposition}
 \label{P21IX11.1}
 Let $\Omega$ be a conditionally compact domain with Lipschitz boundary and let $Z$ be a vector field on an $(n+1)$-dimensional manifold $\mcM$ with continuous metric $g$ and metric volume element $d\mu_g= \sqrt{|\det g|}d^ {n+1}x$. If
 $$
  Z\in W^{1,1}_\loc(\mcM)\;,
   \quad
   g\in W^{1,n}_\loc(\mcM)
  \;,
 $$
 then the Stokes identity holds:
\bel{21IX12.1}
 \int_{\partial \Omega}  Z \rfloor d\mu_g =  \int_{\Omega} \nabla_\alpha Z^ \alpha \, d \mu_g
  \;.
 \ee
\end{Proposition}

\proof
If  the metric and $Z$ are Lipschitz, then \eq{21IX12.1} is standard (see, e.g., \cite{DriverNotes,Mattila,FedererMeasureTheory}).  By density, it remains to show that both the left-hand side and the right-hand side are continuous in the topologies indicated.

We start with the volume integral which, in local coordinates, takes the form
\bel{21IX12.2}
  \int  \nabla_\alpha Z^ \alpha d \mu_g
  =
  \int  \left( \partial_\alpha Z^ \alpha + \Gamma^\alpha_{\alpha\beta} Z^\beta \right) \sqrt{|\det g|} \,d ^{n+1} x
  \;.
\ee
The $\partial Z$ term is clearly continuous. Concerning the $\Gamma Z$ terms, it suffices to use $\Gamma \in L^n _\loc(\mcM)$ together with the continuous Sobolev embedding $W^{1,1}_\loc(\mcM) \subset L^{\frac n {n-1}}_\loc (\mcM)$ to obtain the desired conclusion.

Consider, next,   the boundary integral in \eq{21IX12.1}. The result would be standard using trace theorems if $\partial \Omega$ were smooth. This is not the case, so an argument is needed: By definition of a Lipschitz hypersurface, for every $p\in \partial \Omega$ there exists local coordinates $x^\mu$  near $p $ in which  $\partial \Omega$ takes the form
$$
 x^0 = f(x^i)
 \;,
$$
where $f$ is Lipschitz continuous. Let   new coordinates $y^\mu$ be defined as
$$
 y ^0 = x^0- f(x^i)\;,
 \quad
 y^i = x^i
 \;.
$$
Set
$$
 \bar Z ^\alpha ( y) = Z^\alpha ( x(y))
 \;.
$$
The map $x\to y$ is  Lipschitz, with unit Jacobian wherever defined, and by the change-of-variables theorem for Lipschitz maps we find, for each $\alpha$,
\beaa
\lefteqn{
  \int \big(|Z^\alpha| + \sum_ \mu \big | \partial _{x^\mu} Z^\alpha\big| \big) d^{n+1} x
  }
  &&
\\
  && =
  \int \big(|\bar Z^\alpha| + \big | \partial _{y^0} \bar Z^\alpha\big|
  +\sum_ i \big | \partial _{y^i} \bar Z^\alpha - \frac{\partial f}{\partial x^i} \partial _{y^0} \bar Z^\alpha\big|  \big) d^{n+1} y
  \;.
\eeaa
This implies that $\bar Z \in W^{1,1}_\loc(\R^{n+1})$. In the coordinates $y$ the boundary $\partial \Omega$ reads $\{y^0=0\}$, so the usual trace theorems apply to give
$\bar Z|_{\partial \Omega} \in W^{1/2,1}_\loc \subset L^1 (\partial \Omega)$. From
\bel{21IX12.3}
 \int_{y^0=0}  Z \rfloor d\mu_g =  \int_{y^0=0} \left(\bar Z^ 0 - \frac {\partial f}{\partial x^i} \bar Z ^i\right)\sqrt{|\det g|} d ^{n } y
 \ee
 one easily infers that the boundary-integral is continuous in the $  W^{1,1}_\loc(\mcM)$ topology for $Z$ and in the topology of locally-uniform convergence for $g$.
\qedskip

The following corollary of Proposition~\ref{P21IX11.1} is of main interest to us:

\begin{Corollary}
 \label{C21IX12.1}
 Suppose that $g$ is a continuous metric and assume that
 $$  g\in W^{1,n}_\loc(\mcM)\;, \quad
 \psi= (\psi^A)\in H^2_\loc(\mcM)
   \;.
$$
   Then Stokes' theorem applies to the ``current vector field"
   \bel{6IV13.2}
   Z_\mu \equiv T_{\mu \nu} X^\mu := \sum_A \big(\partial_\mu \psi^A  \partial_\nu \psi^A  -
 \frac 12 g^{\sigma\rho} \partial_\sigma \psi^A \partial_ \rho \psi^A g_{\mu\nu} + \psi^A \psi ^A g_{\mu\nu}
  \big) X^\mu
    \;,
   \ee
   where
    $X$ is any locally Lipschitz
vector field.
\end{Corollary}

\proof
We have, symbolically, $Z \approx g \psi^2 +g^{-1} (\partial \psi)^2$, which is clearly in $L^1_\loc(\mcM)$. Further,
$$
 \partial Z \approx \partial g \psi^2 + g \psi \partial \psi + g^{-1} \partial g (\partial \psi)^2  +  g^{-1}  \partial \psi   \partial^2 \psi
 \;.
$$
The second and last terms are obviously in $L^1_\loc(\mcM)$ for $\psi$'s which are in $H^2_\loc(\mcM)$. For the before last, we have $\partial g\in L^n_\loc (\mcM)$, while by Sobolev embedding it holds that $\partial \psi \in L^{\frac{2n}{n-1}}_\loc(\mcM)$, and the fact that this term is also in $  L^1 _\loc(\mcM)$ readily follows.
The analysis of the first   term  is similar to that of the third. Thus the hypotheses of Proposition~\ref{P21IX11.1} are satisfied.
\qed

\begin{Remark}
 \label{R6IV13.1}
Consider a collection of fields $\varphi^I$ satisfying a semi-linear system of wave-equations which, in local coordinates, take the form
\bel{6IV13.1}
 g^{\mu\nu}\partial_\mu\partial_\nu\varphi^I  = \chi^I (  \partial_\mu \varphi^J,  \varphi^J,x)
   \;,
\ee
for some functions $\chi^I$.
For $k\ge 1$ the $k$-th energy inequality for the fields $\varphi^I$ can be derived using the current vector field $Z$ defined in \eq{6IV13.2} with
$$
 (\psi^A) = \left(\frac{\partial^j \varphi^I}{\partial x^{\alpha_1} \cdots \partial x^{\alpha_j}}\right)_{0\le j\le k-1}
  \;.
$$
For example, for the Einstein equations in a coordinate-gauge as in \eq{6IV13.3}, this gives local-in-time control of the $\Hksp$ semi-norms of the metric by taking $(\varphi ^I) = (g_{\mu\nu})$.
\end{Remark}

\syncx{file SpaceSobolevManifolds; a longer version but with problems in James Grant's directory as SpaceSobolevManifoldsOrig}
\section{Manifolds of $W^{k+1,p}_\myspaceloc $ differentiability class}
\label{splocAppAGC}

 \syncx{needs synchronizing with the Sobolev manifolds section, and checking for duplication}
When using wave-coordinates for a $\Hssp$ metric one needs to work with coordinate transformations which are not smooth but of $\Hspsp$ differentiability class. This begs the question, what happens with functions and tensor fields under such coordinate changes. This is the main issue addressed in this appendix.

Consider a smooth manifold $\mcM $; on such a manifold one can define in an invariant way tensor
fields which are of $C^\infty$ differentiability class, similarly for   $C^k$, $W^{k,p}_\loc $,   or   $W^{k,p}_\myspaceloc $ class. For example, one says that a tensor
field is of $W^{k,p}_\myspaceloc $ class if there exists a covering of $\mcM $ by
coordinate patches such that the coordinate components of the tensor
in question are in $W^{k,p}_\myspaceloc $ in each of the coordinate patches.
Now, even though the transition functions when going from one coordinate system
to another are smooth, it is not clear that  $W^{k,p}_\myspaceloc $-differentiability will be true in all coordinate
system, because the integral differentiability conditions might fail to hold on the constant-$x^0$-slices of some  smooth coordinate systems, unless some preferred coordinate $x^0$ has been chosen.
To address this issue for the metric tensor, we will  assume for definiteness wave coordinates and vacuum field equations. In fact, the  invariance of the $W^{k,p}_\myspaceloc $-differentiability conditions holds for any tensor field satisfying wave-type equations, and many alternative coordinate-conditions for vacuum metrics are possible as e.g.\ in \eq{6IV13.3}. Note that this problem does not arise for $C^\infty$,   $C^k$, or $W^{k,p}_\loc $ tensor fields.

Let us start with a definition:

\begin{Definition}
 \label{D5IV13.1} Let $k\in \R$, $p\in [1,\infty]$.
 A function  $f$ will be said to be of $ W^{k,p}_\spaceloc $-differentiability class if every point $p\in \mcM$ has a coordinate neighborhood $\mcU =I\times \mcV $, where $I$ is the range of a  coordinate $t\equiv x^0$, with the following properties:
 On every level set $\hyp_\tau\subset \mcU $ of the coordinate $t$ we have  $f\in W^{k,p}_\loc (\hyp_\tau)$, with the time-derivatives of order $0\le \ell \le k$ of $f$ being of $ W^{k-\ell,p}_\spaceloc $ differentiability. Furthermore the functions
$$
 I \ni t \mapsto \|\partial^\ell_t f |_{\hyp_t}\|_{W^{k-\ell,p}}
  \;,
 \qquad 0\le \ell \le  k
$$
are required to be continuous.

We set $\Hksp:=W^{k,2}_\spaceloc $.
\end{Definition}

To continue, it is convenient to introduce the following notation: for $x,y\in   \R\cup\{\infty\}$,
we will write $x\gst y $ if the following holds:
\begin{equation}
  \label{sploceq:lst}
  x\gst y \Longleftrightarrow \cases{x\ge y \ ,&  if $y>0$\ , \cr
                                     x> y  \ , & if $y\le0$\  .}
\end{equation}
(Note that for $x\ge0$ the only value of $x$ at which ``$\gst$''
does not coincide with ``$\ge$'' is $x=0$.)  In this notation
the Sobolev embedding theorem, in dimension $n$, can be stated as~\cite{adams:sobolev:spaces,Aubin}:
\begin{equation}
  \label{sploceq:sobemb}
  W^{s,t}_\myspaceloc   \subset W^{u,v}_\myspaceloc  \quad  \Longleftrightarrow\quad u\le s
\ \mbox{  and  }\ \frac{1}{v} \gst \frac{1}{t}-\frac{s-u}{n}\ \;.
\end{equation}

We have the following:

\begin{Lemma}
  \label{sploclcompCl}
  Let $\Omega$ and $ \cU $ be open subsets of $ \R^{n+1}$,  and let $\Psi: \Omega\to \cU $ be a
  $C^1$ diffeomorphism such that $\Psi\in \Wkppsp(\Omega;\bR^n)$ (with respect to the coordinate $x^0$),
   \syncx{should be defined}
   with
   $p\in [1,\infty]$, $k\in \N$, $kp>n$.
   Write $\Psi=\big(\Psi^0  (x^0,x^k),\Psi^\ell  (x^0,x^k)\big)$, and let $\ol\hyp_\tau$ denote the level sets of $\Psi^0$, assume that $\partial_0\Psi^0$ has no zeros so that the $\ol\hyp_\tau$ are graphs which can be parametrized by $x^i$.
  Let $(\ell,q)$ be such that the $n$-dimensional Sobolev embedding
  $\Wkppsp\subset\wlqgl$
  holds. If for every $\tau$ such that $\ol \hyp_\tau \ne \emptyset$, for every compact $K$ satisfying $K\subset \ol \hyp_\tau$, and for every $0\le j\le \ell$ we have
  \bel{5IV13.1}
   \int_K \bigg|\frac{\partial ^j F}{\partial x^{\alpha_1}\cdots \partial x^{\alpha_j}}\bigg|^q d^n x <\infty
   \;,
  \ee
  then
$$F\circ \Psi \in 
\Wlqsp(\Omega)\ .$$
\end{Lemma}

\proof
The proof is a straightforward adaptation of that of Lemma~A.2 in~\cite{BartnikChrusciel1}.
\qedskip

We also need:

\begin{Lemma}
  \label{sploclproductCl}
  Let $0\le m\le \ell \le k$, $q,p\in [1,\infty]$, $kp>n$. Suppose
  that $(\ell,q)$ is such that the $n$-dimensional Sobolev embedding $\Wkpsp \subset
  W^{\ell,q}_\myspaceloc $ holds. Then the product map
  %
%
$$
\Wkmpsp \times \Wlqsp \ \ni \ (f,g) \ \longrightarrow
fg \in \Wlmqsp
$$
is continuous.
\end{Lemma}

\proof
The proof is a repetition of that of Lemma~A.4 in~\cite{BartnikChrusciel1}.
\qedskip

Finally:

\begin{Proposition}
  \label{splocpinvnew}  Under the conditions of Lemma~\ref{sploclcompCl}, let $F$ be a tensor field the cooordinate-components of which satisfy \eq{5IV13.1}. If moreover   the $n$-dimensional Sobolev embedding
    $$
    \Wkpsp \subset W^{\ell,q}_\myspaceloc
    $$
    holds, then $\Psi^* F \in W^{\ell,q}_\myspaceloc $.
\end{Proposition}

\proof This follows from Lemmata~\ref{sploclcompCl}-\ref{sploclproductCl}, as in~\cite[Appendix~A]{BartnikChrusciel1}.
\qedskip

\bibliographystyle{amsplain}
\bibliography{../references/hip_bib,%
../references/reffile,%
../references/newbiblio,%
../references/newbiblio2,%
../references/chrusciel,%
../references/bibl,%
../references/howard,%
../references/bartnik,%
../references/myGR,%
../references/newbib,%
../references/Energy,%
../references/dp-BAMS,%
../references/prop2,%
../references/besse2,%
../references/netbiblio}

\end{document}